\documentclass{ar2e}
\usepackage{ulem}  
\usepackage{color} 
\usepackage{graphicx} 
\newcommand{\singlespace}{\tiny\renewcommand{\baselinestretch}{1.0}\normalsize}

\begin{document}

\input epsf.def
\input psfig.sty

\jname{Annu. Rev. Phys. Chem.}
\jyear{2011}
\jvol{1}
\ARinfo{1056-8700/97/0610-00}

\singlespace

\title{{\color{red} Progress in Time-Dependent Density-Functional Theory}}

\markboth{Casida \& Huix-Rotllant}{Progress in Time-Dependent Density-Functional Theory}

\author{M. E. Casida and M. Huix-Rotllant
\affiliation{Laboratoire de Chimie Th\'eorique, 
             D\'epartement de Chimie Mol\'eculaire (DCM, UMR CNRS/UJF 5250), 
             Institut de Chimie Mol\'eculaire de Grenoble (ICMG,FR2607), 
             Universit\'e Joseph Fourier (Grenoble I), 
             301 rue de la Chimie, BP 53, 
             F-38041 Grenoble Cedex 9, France. ; 
             email: mark.casida@ujf-grenoble.fr, miquel.huix@ujf-grenoble.fr}}


\begin{abstract}

The classic density-functional theory (DFT) formalism introduced by Hohenberg, Kohn, and Sham 
in the mid-1960s, is based upon the idea that the complicated $N$-electron wavefunction can be
replaced with the mathematically simpler $1$-electron charge density in electronic structure
calculations of the ground stationary state.  As such, ordinary DFT is neither able to treat 
time-dependent (TD) problems nor describe excited electronic states.  In 1984, Runge and Gross
proved a theorem making TD-DFT  formally exact.  Information about
electronic excited states may be obtained from this theory through the linear response (LR)
theory formalism.  Beginning in the mid-1990s, LR-TD-DFT became increasingly popular for
calculating absorption and other spectra of medium- and large-sized molecules.  Its ease
of use and relatively good accuracy has now brought LR-TD-DFT to the forefront for this type
of application.  As the number and the diversity of applications of TD-DFT has grown, so too 
has grown our understanding of the strengths and weaknesses of the approximate functionals
commonly used for TD-DFT.  The objective of this article is to continue where a previous review of
TD-DFT in this series [Annu.\ Rev.\ Phys.\ Chem.\ 55: 427 (2004)] left off and highlight
some of the problems and solutions from the point of view of applied physical chemistry.  Since
doubly-excited states have a particularly important role to play in bond dissociation and
formation in both thermal and photochemistry, particular emphasis will be placed upon 
the problem of going beyond or around the TD-DFT adiabatic approximation which limits TD-DFT calculations 
to nominally singly-excited states. \newline 

{Posted with permission from the Annual Review of Physical Chemistry, Volume 63 \copyright 2012  by Annual Reviews, {http://www.annualreviews.org}.}

\end{abstract}

\maketitle
\section{{\color{red} INTRODUCTION}}
\label{sec:intro}

\marginpar{\hrule  $\,$ \\ Space and spin coordinates have been denoted by, $1,2,\cdots={\bf r}_1 \sigma_1, {\bf r}_2\sigma_2,\cdots$.  \\ \hrule}
Even after the Born-Oppenheimer separation, the time-independent Schr\"odinger equation for
$N$ electrons 
in an ``external potential'' consisting of $M$ nuclei and any applied electric fields,
$\hat{H} \Psi(1,2,\cdots,N) = E \Psi(1,2,\cdots,N)$, is still notoriously difficult to solve.   
\marginpar{\hrule $\,$ \\ 
{\color{blue} DFT}: density-functional theory.  A {\color{blue} functional} is a function of a function.  It is designated by
a square bracket notation.  Thus $v[\rho]({\bf r})$ is simultaneously a functional of the function $\rho$
and a function of the position, ${\bf r}$.  \\ \hrule}
Yet such amazing progress has been made in solving this equation for increasingly complex systems 
since its introduction early in that century that we may well refer to the 20th century as the 
``Century of Quantum Mechanics.''  Of course, computers have helped quite a bit, but improved 
approximation methods based upon progress in our understanding of the mathematical and physical 
properties of the underlying objects of quantum mechanics has also been instrumental in finding more 
efficient solutions.  
One of the landmarks has been Hohenberg-Kohn-Sham \cite{HK64,KS65} density-functional
theory (DFT), in which it is recognized that the complicated many-electron wave function, $\Psi(1,2,\cdots,N)$,
may be replaced with functionals of a simpler object---namely the charge density, 
$\rho({\bf r}_1) = N \sum_{\sigma_1} \int \int \cdots \int \vert \Psi({\bf r}_1 \sigma_1 ,2,\cdots,N) \vert^2 
\, d2 d3 \cdots dN$--- in calculating properties of the ground-stationary state.  
\marginpar{\hrule  $\,$ \\ Unless explicitly present, we take $\hbar=m_e=e=1$ (i.e.,
we will use Hartree atomic units) throughout this article. \\ \hrule}
DFT was a mighty advance but it was to remain pretty much in the shadow of wave-function theory as far as 
quantum chemistry was concerned until into the 1980s when the tide turned.
Approximate exchange-correlation functionals had advanced to the point where they could often compete
with accurate {\it ab initio} wave-function methods, particularly for larger molecules of experimental
interest.  Nevertheless the restriction to the ground-stationary state was clearly disappointing 
\marginpar{\hrule  $\,$ \\ 
{\color{blue} TD}: time dependent \\
{\color{blue} UV-Vis}: ultraviolet-visible\\
{\color{blue} NLO}: nonlinear optics
\\ \hrule}
for such popular applications as UV-Vis spectroscopy, nonlinear optics (NLO),
and photochemistry.  
But, as with everything else, it seemed that ``it was all just a matter of time.''
The formal basis of modern time-dependent (TD) DFT is the Runge-Gross theorem showing that,
for a system initially in its ground stationary state exposed to a time-dependent perturbation,
the time-dependent charge density, $\rho({\bf r},t)$, determines the time-dependent external potential
up to an additive function of time \cite{RG84}.
\marginpar{\hrule  $\,$ \\ 
{\color{blue} {\it ab initio}}: Latin meaning ``from the beginning'' used in
quantum chemistry for first principles wave function calculations but often includes DFT in other contexts\\ 
{\color{blue} LR}: linear response
\\ \hrule}
Once again there is a formally-exact functional which must be approximated but which then gives immediate
access to NLO properties and, via linear-response (LR) theory, to excited-state information.
The LR-TD-DFT equations introduced by one of us \cite{C95} have been programmed in most quantum chemistry
and quantum physics electronic structure codes and are probably currently the mostly widely-used method to
treat the excited states of medium- to large-sized molecules.
Many review articles have been written on various aspects of TD-DFT,
some special volumes devoted to TD-DFT have appeared or will soon appear, and
three books have been published or will soon be published 
(see the references cited in Ref.~\cite{C09} as well as the newer Refs.~\cite{BKK11,GMNR11}.)
The present review may 
be regarded as a follow up to an earlier review of TD-DFT in the present series \cite{MG04}.
\marginpar{\hrule $\,$ \\
{\color{blue} Ref.~\cite{MG04}}:
Marques MAL, Gross EKU, Annu.\ Rev.\ Phys.\ Chem.\ 55:427 (2004), ``Time-Dependent Density-Functional Theory''.
\\ \hrule}

\begin{figure}
  \begin{center}
    \includegraphics[angle=0,width=1.0\textwidth]{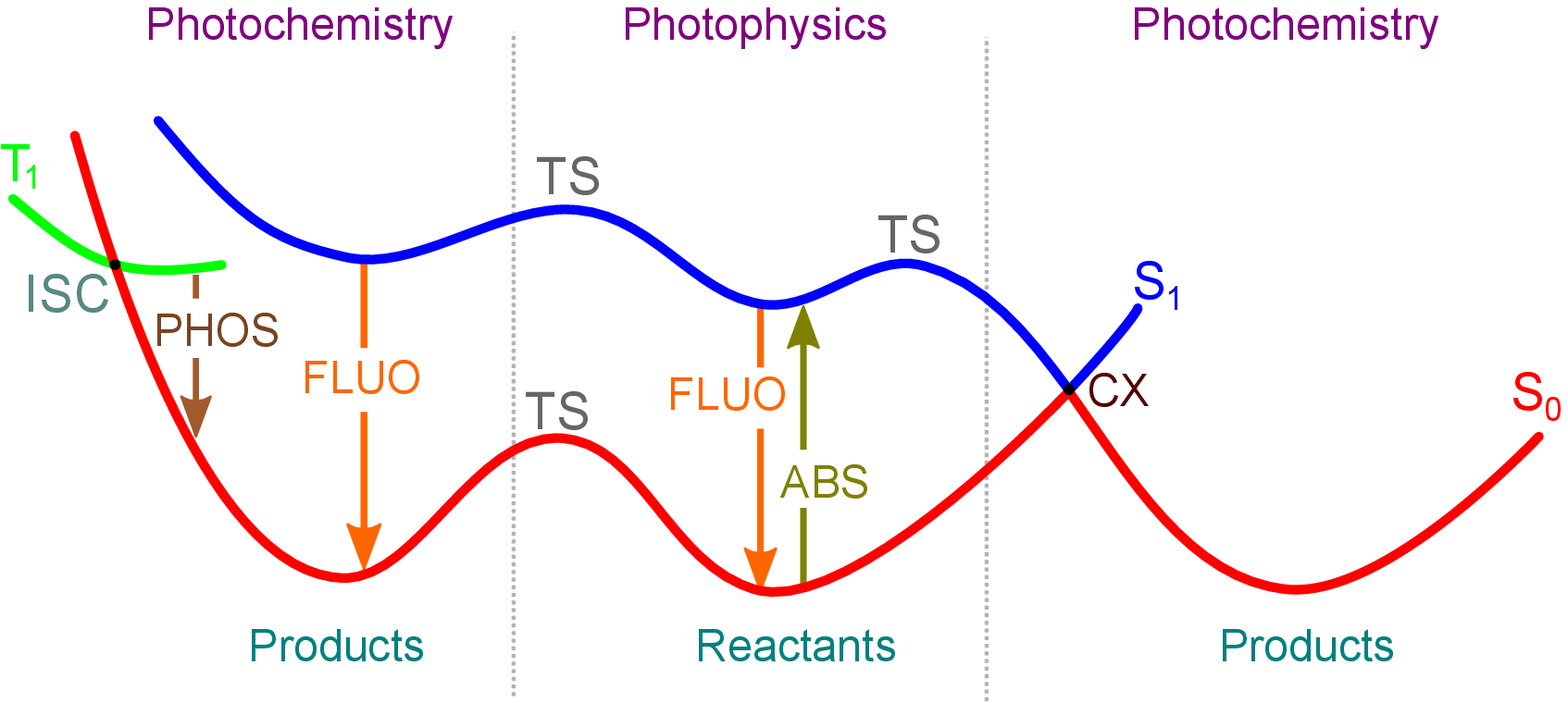}
  \end{center}
  \caption{\label{fig:PES}
    Schematic representation of potential energy surfaces for photophysical and photochemical
    processes: $S_0$, ground singlet state; $S_1$, lowest excited singlet state; $T_1$, lowest
    triplet state; {\color{blue} ABS}, absorption; {\color{blue} FLUO}, fluorescence; 
    {\color{blue} PHOS}, phosphorescence; {\color{blue} ISX}, intersystem crossing; {\color{blue} CX}, 
    conical intersection; {\color{blue} TS}, transition state.}
\end{figure}

We take the point of view of theoretical physical chemists interested in spectroscopy and photoprocesses and
particular emphasis will be given to advances made over the last seven years.  
{\bf Figure~\ref{fig:PES}} provides a schematic summary, sampling the breadth of chemical and physical phenomena
that physical chemists might wish to describe.  We will often want to refer to various features on this diagram.
We will also often take pains to first review wave-function theory before discussing (TD-)DFT.  Our idea in
doing this is to concentrate more fully afterwards on what is different in (TD-)DFT when compared to 
wave-function theory.

This review is organized as follows.  The next section reviews elementary wave-function theory.
Section~\ref{sec:dft} reviews conventional time-independent ground state DFT.  This is necessary in order 
to keep the chapter somewhat self-contained because TD-DFT shares many 
features in common with DFT and problems in approximate functionals used in TD-DFT may often be traced back
to problems inherited from approximate functionals in TD-DFT.  Section~\ref{sec:tddft} then introduces formal
\marginpar{\hrule $\,$ \\ {\color{blue} AA}: adiabatic approximation \\ \hrule} 
TD-DFT, the adiabatic approximation (AA) which defines conventional TD-DFT, typical applications, and then 
discusses problems with conventional TD-DFT, how to test for problems, and how problems can be fixed or 
at least attenuated.  Some of these problems come from the very nature of the TD-DFT AA.
TD-DFT should be exact provided the exact frequency- and initial state-dependent exchange-correlation (xc) 
functional of TD-DFT is used.
\marginpar{\hrule $\,$ \\ {\color{blue} xc}: exchange-correlation \\ \hrule} 
This is not the same as the xc-functional of ordinary DFT which is simply borrowed by TD-DFT whenever the
AA is used and so a true TD-DFT xc-functional should go beyond the AA
to include frequency and possibly also initial state dependence.  Section~\ref{sec:beyond} discusses 
what is known about the exact xc-functional of TD-DFT and discusses attempts at creating and using approximate
functionals which go beyond the TD-DFT AA.  As these attempts are still in a more or less
embryonic stage, it is also interesting to consider practical working methods to overcome the limitations
of the TD-DFT AA without introducing any frequency dependence.
This is the topic addressed in Sec.~\ref{sec:around}.  Section~\ref{sec:conclude} sums up with some perspectives
regarding future developments in TD-DFT.

\section{{\color{red} WAVE-FUNCTION THEORY}}
\label{sec:wft}

Modern DFT and TD-DFT resemble wave function theory in its simplest incarnations.  For this reason 
it is useful to review basic wave-function theory if only to establish some notation and 
to make sure that key concepts are fresh in the mind of the reader.  

\subsection{\color{red} Hartree-Fock Approximation}

According to basic quantum mechanics, it is the wavefunction, $\Psi$, which describes the fundamental state
of the system and so we should aim to solve the Schr\"odinger equation. Except
for exceptionally simple systems, we can only solve the Schr\"odinger
equation approximately with the help of powerful approximation techniques such as the well-known 
variational principle.

\marginpar{\hrule $\,$ \\ {\color{blue} HF}: Hartree-Fock. \\ \hrule}
The Hartree-Fock (HF) approximation consists of using a trial wavefunction in the form of a single
Slater determinant of orthonormal orbitals.
Varational minimization of the energy subject to the orbital orthonormality constraint leads to the
HF equation,
\begin{equation}
  \hat{f} \psi_i(1) = \epsilon_i \psi_i(1) \, , \label{eq:wft.3}
\end{equation}
given in a spin-orbital notation.
Here the one-electron Fock operator, $\hat{f}$, is the sum of the one-electron kinetic energy operator, $\hat{t}$,
the external potential, $v_{ext}({\bf r})$, which is a multiplicative operator describing the interaction of the
electron with the electric fields generated by the $M$ nuclei and any applied electronic fields.  
\marginpar{\hrule $\,$ \\ {\color{blue} SCF}: Self-consistent field. \\ \hrule}
The Fock operator also includes a self-consistent field (SCF) composed of a Hartree term, 
$v_H({\bf r}_1) = \int \rho({\bf r}_2)/r_{12} \, d{\bf r}_2$,
(also known as the Coulomb term), and an exchange operator, $\hat{\Sigma}_x$ (designated here 
as the exchange self-energy using the terminology and notation of Green's function theory.)
As described in elementary quantum chemistry textbooks, 
\begin{equation}
  \hat{\Sigma}_x \phi(1) = - \int \frac{\gamma(1,2)}{r_{12}} \phi(2) \, d2 \, ,
  \label{eq:wft.4}
\end{equation}
is a complicated integral operator, where $\gamma(1,2) = \sum_i \psi_i(1) n_i \psi_i^*(2)$ is the one-electron
reduced density matrix (1-RDM) and the $n_i$ are orbital occupation numbers.
\marginpar{\hrule $\,$ \\ {\color{blue} 1-RDM}: one-electron reduced density matrix. \\ \hrule}
For completeness, we note the HF energy expression for the total energy,
\begin{equation}
  E = \sum_i n_i \langle \phi_i \vert \hat{t} \vert \phi_i \rangle + \int v_{ext}({\bf r}) \rho({\bf r}) \, d{\bf r}
    + E_H[\rho] + E_x[\gamma] \, , \label{eq:wft.5}
\end{equation}
where $\hat{t}$ is the one-electron kinetic energy operator, $v_{ext}$ is any potential external (i.e., the nuclear
attraction and any applied electric fields) external to the system of $N$-electrons,
the Hartree energy,
\begin{equation}
   E_H[\rho] = \frac{1}{2} \int \int \frac{\rho({\bf r}_1)\rho({\bf r}_2)}{r_{12}} \, d{\bf r}_1 d{\bf r}_2 \, ,
   \label{eq:wft.6}
\end{equation}
and the exchange energy,
\begin{equation}
   E_x[\gamma] = -\frac{1}{2} \int \int \frac{\vert \gamma(1,2) \vert^2}{r_{12}} \, d1 d2 \, .
   \label{eq:wft.7}
\end{equation}
According to Koopmans' theorem \cite{K34}, the occupied orbital energies, $\epsilon_i$, in Eq.~(\ref{eq:wft.3})
may be interpretted as minus ionization potentials and the unoccupied orbital energies as minus electron affinities.
That is, occupied orbitals ``see'' $N-1$ electrons while unoccupied orbitals ``see'' $N$ electrons.

\subsection{\color{red} Excited States}

Let us now turn to the excited-state problem.  We will begin with the equation-of-motion (EOM) formalism.  
This has the advantage of conceptual simplicity and, though we will not show it here, can be directly 
linked to linear response theory \cite{JS81}.
The operator, $\vert I \rangle \langle 0 \vert$, which destroys the ground state ($\vert 0 \rangle$)
and creates the $I$th excited state ($\vert I \rangle$) and its adjoint, $\vert 0 \rangle \langle I \vert$,  are solutions 
of the EOM, 
\begin{equation}
   [\hat{H},\hat{\cal O}^\dagger] = \omega \hat{\cal O}^\dagger \, .  
   \label{eq:wft.8}
\end{equation}
Evidently $\omega=E_I-E_0$ is an excitation 
energy in the case of $\vert I \rangle \langle 0 \vert$ and $\omega=E_0-E_I$ is a de-excitation in the case of 
$\vert 0 \rangle \langle I \vert$.

\marginpar{\hrule $\,$ \\ {\color{blue} EOM}: equation-of-motion.\\
{\color{blue} $n$p$m$h}: $n$-particle/$m$-hole.\\ \hrule}
We will now seek a solution of the EOM of the form, 
\begin{equation}
  \hat{\cal O}^\dagger = \sum_{i,a} a^\dagger i X_{ia} + \sum_{i,a} i^\dagger a Y_{ia} \, .
  \label{eq:wft.9}
\end{equation}
That is we develop $\hat{\cal O}^\dagger$ in a basis set of one-particle/one-hole (1p1h) excitation and
de-excitation operators.  This operator basis and the EOM operator ``metric'' defined by
$(\hat{\cal A},\hat{\cal B})=\langle \Phi_{HF} \vert \left[ \hat{\cal A}^\dagger,
\hat{\cal B} \right] \vert \Phi_{HF} \rangle$ then allows us to develop a matrix form of the EOM, namely,
\begin{equation}
  \left[ \begin{array}{cc} {\bf A} & {\bf B} \\ {\bf B}^* & {\bf A}^* \end{array} \right]
  \left( \begin{array}{c}  {\vec X} \\  {\vec Y} \end{array} \right) 
   = \omega \left[ \begin{array}{cc} {\bf 1} & {\bf 0} \\ {\bf 0} & -{\bf 1} \end{array} \right]
  \left( \begin{array}{c}  {\vec X} \\ {\vec Y} \end{array} \right) \, ,
  \label{eq:wft.10}
\end{equation}
where the notation ${\vec X}$ and ${\vec Y}$ is deliberately chosen to indicate that these objects are to be treated as
column vectors in Eq.~(\ref{eq:wft.10}).  The matrices are defined by,
\marginpar{\hrule $\,$ \\ {\color{blue} MO}: molecular orbital.\\ \hrule}
\begin{eqnarray}
  A_{ia,jb} & = & \delta_{i,j} \delta_{a,b} \left( \epsilon_a - \epsilon_i \right) + K_{ia,jb} \nonumber \\
  B_{ia,jb} & = & K_{ia,bj} \, ,
  \label{eq:wft.11}
\end{eqnarray}
where the coupling matrix,
\begin{equation}
  K_{pq,rs} = (pq \vert f_H \vert sr) - (pr \vert f_H \vert sq) \, .
  \label{eq:wft.12}
\end{equation}
Note that here and hence forth we adapt the molecular orbital (MO) index convention,
\marginpar{\hrule $\,$ \\ {\color{blue} RPA}: random phase approximation.
Sometimes one talks of RPA with exchange ({\color{blue} RPAE} or {\color{blue} RPAX}) in order to 
emphasize inclusion of the second integral in Eq.~(\ref{eq:wft.12}).
However it is often left up to the reader to figure out whether this integral is included or not in the term RPA.
\\ \hrule}
\begin{equation}
  \underbrace{a,b,c,\cdots,g,h}_{unoccupied} \underbrace{i,j,k,l,m,n}_{occupied} \underbrace{o,p,q,\cdots,y,z}_{free} \, ,
  \label{eq:wft.13}
\end{equation}
where ``free'' means ``free to be either occupied or unoccupied.''  
We are also using a ``lazy version'' of second-quantized notation where $p=\hat{a}_p$ and $p^\dagger = \hat{a}_p^\dagger$,
and Mulliken ``charge cloud'' notation,
\begin{equation}
  ( pq \vert f \vert s r ) = \int \int \psi_p^*(1) \psi_q(1) f(1,2) \psi_s^*(2) \psi_r(2) \, d1 d2 \, ,
  \label{eq:wft.14}
\end{equation}
where we have taken $f$ to be the Hartree kernel, $f_H(1,2) = 1/r_{12}$.

Equation~(\ref{eq:wft.10}) is often called the random phase approximation (RPA) for purely historical reasons.
Some caution is in order when
using this term since the RPA was originally developed in nuclear and solid-state physics without the second integral 
in the coupling matrix [Eq.~(\ref{eq:wft.12})].  
Although we have chosen an EOM derivation of Eq.~(\ref{eq:wft.12}), this equation also
emerges from LR-TD-HF  and so may be referred to by that name (see for example, Ref.~\cite{JS81} pp.\ 122-134, 144-151). 
It is a pseudoeigenvalue problem with an operator ``overlap matrix'' on the right-hand side which is simplectic
rather than positive definite as would be a true overlap matrix.  It also has paired excitation, $(\vec{X},\vec{Y})$,
and de-excitation solutions, $(\vec{Y}^*,\vec{X}^*)$, whose corresponding excitation/de-excitation energies, $\omega$, differ
only by a sign change.  The normalization of the solutions is given by $(\hat{\cal O}^\dagger \vert \hat{\cal O}^\dagger )
= \vec{X}^\dagger \vec{X} - \vec{Y}^\dagger \vec{Y}$, which is typically positive for excitations and negative for
de-excitations.  
\marginpar{\hrule $\,$ \\ {\color{blue} PES}: potential energy surface. \\ \hrule}
Potential energy surfaces (PESs) may be calculated from the excited-state energies, $E_I = E_{HF}+\omega_I$.
Since analytic derivatives are available for LR-TD-HF \cite{O94}, automatic searching for critical points on excited-state 
PESs ({\bf Fig.~\ref{fig:PES}}) is possible in this model.

An important point that will only be mentioned here is that the LR-TD-HF equation provides not only excitation energies,
$\omega_I$, but also the corresponding oscillator strengths,
\begin{equation}
  f_I =  \frac{2m_e}{3\hbar} \omega_I \vert \langle 0 \vert {\bf r} \vert I \rangle \vert^2
  \label{eq:wft.15}
\end{equation}
which may be calculated from the coefficient vector, $(\vec{X},\vec{Y})$, and the dipole matrix.
\marginpar{\hrule $\,$ \\ {\color{blue} TRK}: Thomas-Reiche-Kuhn $f$-sum rule. \\ \hrule}
Oscillator strengths are pure numbers and, in a complete basis set (or other basis set satisfying the relation
$\left[ x, \hat{p}_x \right] = +i\hbar$, satisfy the Thomas-Reiche-Kuhn (TRK) $f$-sum rule,
$\sum_I f_I = N$, where $N$ is the number of electrons.  Equation~(\ref{eq:wft.15}) has  been deliberately expressed in 
Gaussian, rather than atomic, units so that it may be more easily related to the molar extinction coefficient, $\epsilon$,
in an absorption experiment.  
\marginpar{\hrule $\,$ \\ {\color{blue} SI}: {\it Syst\`eme International}. \\ \hrule}
In SI units, the frequency spectrum is given to
a first approximation by,
\begin{eqnarray}
  \epsilon(\nu) & = & \sum_I \frac{N_A e^2}{4 m_e c \ln(10) \epsilon_0} {\cal S}(\nu-\nu_I) \nonumber \\
                & = & \left(\mbox{6.94 $\times$ 10$^{+18}$ L/cm.s}\right) \sum_I {\cal S}(\nu-\nu_I) \, ,
  \label{eq:wft.16}
\end{eqnarray}
where $N_A$ is Avogadro's number and ${\cal S}(\nu)$ is a spectral shape function (typically a Gaussian whose area is
normalized to unity and whose full-width-at-half-maximum is determined from experiment.) 

\marginpar{\hrule $\,$ \\ {\color{blue} TDA}: Tamm-Dancoff approximation. \\ 
{\color{blue} CIS}: configuration interaction singles.\\ \hrule}
An approximation which is frequently made is the Tamm-Dancoff approximation (TDA), which consists of neglecting 
the ${\bf B}$ matrix in the LR-TD-HF equation, thus decoupling the excitations from the de-excitations.  
The TRK $f$-sum rule is lost when the TDA is made, but the resultant equation is variational in the wave function sense.  
In particular, the minimal acceptable model for the excited states of a 
closed-shell molecule is to take the wave function as a linear combination of 1p1h (singly-excited) HF configurations, 
$\Phi^{CIS}_I = \sum_{i,a} a^\dagger i \Phi_{HF} C_{ia,I} = \hat{\cal O}^\dagger_I \Phi_{HF}$.  Use of the variational
principle to find the coefficients which minimize the energy then yields the equation,
\begin{equation}
  {\bf A} \vec{X}_I  =  \omega_I \vec{X}_I \, ,
  \label{eq:wft.17}
\end{equation}
where $\vec{C}_I$ has been renamed $\vec{X}_I$ and the notations ${\bf A} = {\bf H}-E_{HF}$ and $\omega_I = E_I - E_{HF}$
\marginpar{\hrule $\,$ \\ {\color{blue} TOTEM}: two-electron two-orbital model\\ \hrule}
have been introduced in order to make the correspondence with the TDA of LR-TD-HF clear.
Because of the variational nature of the TDA, for some applications, this ``approximation'' sometimes gives better 
results than the original RPA equation ({\it vide infra}).

\begin{figure}
  \begin{center}
    \includegraphics[angle=0,width=1.0\textwidth]{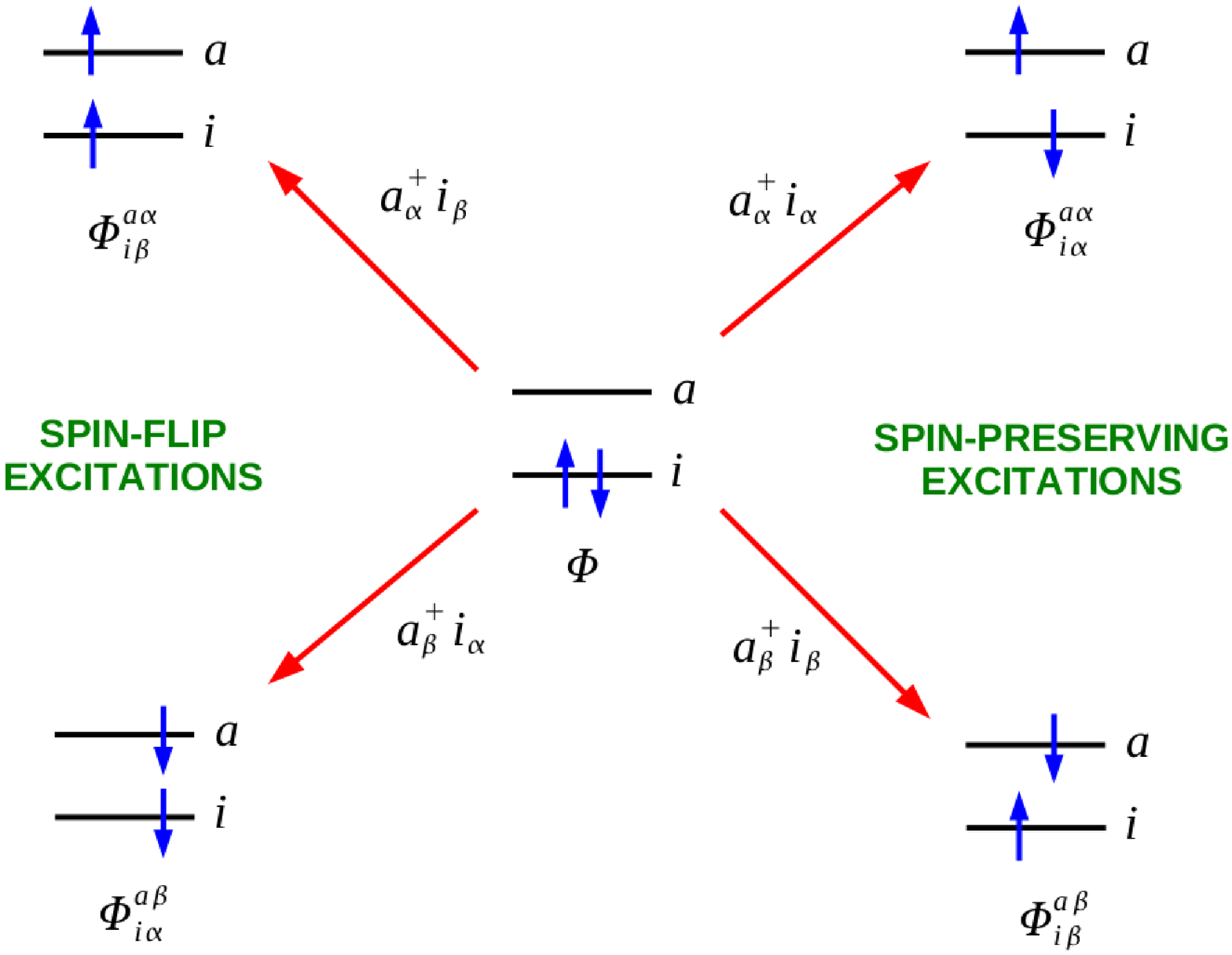}
  \end{center}
  \caption{\label{fig:TOTEM}
    Schematic of excitations in the two-orbital two-electron model. Excitations on the left hand side of the diagram are
    spin-flip excitations while those on the right hand side are spin-preserving excitations.}
\end{figure}

It is useful for interpretational purposes to restrict ourselves to the case where an electron is excited from an
initial MO $i$ to a target MO $a$ neglecting any orbital relaxation.  This two-orbital two-electron model (TOTEM) 
corresponds to the
spin-preserving excitations in {\bf Fig.~\ref{fig:TOTEM}} and yields one singlet and one triplet excitation energy,
as shown in {\bf Table~\ref{tab:TOTEM}}.

\marginpar{\hrule $\,$ \\ {\color{blue} MSM}: multiplet sum method\\ \hrule}
The CIS expressions of {\bf Table~\ref{tab:TOTEM}} may also be derived in a more direct way using the multiplet sum 
method (MSM) as follows:
The wave functions in {\bf Fig.~\ref{fig:TOTEM}} are all eigenfunctions of the spin operator $\hat{S}_z$ but not necessarily
of $\hat{S}^2$.  Linear combinations, $\Phi_{S,M_S}$, are needed to form simultaneous eigenvectors of the two operators,
\begin{eqnarray}
  \Psi_{0,0} & = & \frac{1}{\sqrt{2}} \left( \Phi_{i\alpha}^{a\alpha} + \Phi_{i\beta}^{a\beta} \right) \, \, 
  , \, \,
  \Psi_{1,-1}  =  \Phi_{i\alpha}^{a\beta} \nonumber \\
  \Psi_{1,0} & = & \frac{1}{\sqrt{2}} \left( \Phi_{i\alpha}^{a\alpha} - \Phi_{i\beta}^{a\beta} \right)  \, \, 
  , \, \,
  \Psi_{1,+1}  =  \Phi_{i\beta}^{a\alpha} \, ,
  \label{eq:wft.19}
\end{eqnarray}
where $\Phi_{1,-1}$, $\Phi_{1,0}$, and $\Phi_{1,+1}$ are degenerate triplet states and $\Phi_{0,0}$ is an open-shell
singlet state.  The triplet energy is easily expressed as the expectation value of a single determinant,
\begin{equation}
   E_T = \langle \Psi_{1,-1} \vert \hat{H} \vert \Psi_{1,-1} \rangle 
       = \langle \Phi_{i\alpha}^{a\beta} \vert \hat{H} \vert \Phi_{i\alpha}^{a\beta} \rangle 
       = E[\Phi_{i\alpha}^{a\beta}] \, .
   \label{eq:wft.20}
\end{equation}
The MSM allows the singlet energy, $E_S$, also to be expressed in terms of expectation values of single determinant
wave functions.  To do so, we note that,
\begin{eqnarray}
  E_S & = & \langle \Psi_{0,0} \vert \hat{H} \vert \Psi_{0,0} \rangle
        = \langle \Phi_{i\alpha}^{a\alpha} \vert \hat{H} \vert \Phi_{i\alpha}^{a\alpha} \rangle
          + \langle \Phi_{i\alpha}^{a\alpha} \vert \hat{H} \vert \Phi_{i\beta}^{a\beta} \rangle \nonumber \\
  E_T & = & \langle \Psi_{1,0} \vert \hat{H} \vert \Psi_{1,0} \rangle
        = \langle \Phi_{i\alpha}^{a\alpha} \vert \hat{H} \vert \Phi_{i\alpha}^{a\alpha} \rangle
          - \langle \Phi_{i\alpha}^{a\alpha} \vert \hat{H} \vert \Phi_{i\beta}^{a\beta} \rangle \, .
  \label{eq:wft.21}
\end{eqnarray}
Eliminating the cross term and using Eq.~(\ref{eq:wft.20}) gives us finally,
\begin{eqnarray}
  E_S & = & 2 \langle \Phi_{i\alpha}^{a\alpha} \vert \hat{H} \vert \Phi_{i\alpha}^{a\alpha} \rangle
      - \langle \Phi_{i\alpha}^{a\beta} \vert \hat{H} \vert \Phi_{i\alpha}^{a\beta} \rangle \nonumber \\
      & = & 2 E[\Phi_{i\alpha}^{a\alpha}] - E[\Phi_{i\alpha}^{a\beta}] \, ,
  \label{eq:wft.22}
\end{eqnarray}
where the last line emphasizes that the singlet energy has now been expressed uniquely in terms of the energies
of single Slater determinants.  Subtracting the ground-state HF energy, $E_{HF}$, and explicitly evaluating the
various single-determinantal energies then leads to the CIS formulae in {\bf Table~\ref{tab:TOTEM}}.

\begin{table}
  \caption{
    Summary of formulae within the two-orbital two-electron model ({\bf Fig.~\ref{fig:TOTEM}}).
    $I_i$ is minus the ionization potential of orbital $i$.  $A_a$ is minus the
    electron affinity of orbital $a$.  $A_a(i^{-1})$ is minus the electron
    affinity of orbital $a$ for the ion formed by removing an electron from
    orbital $i$.  $\omega_T$, $\omega_S$, and $\omega_M$ are, respectively,
    $i \rightarrow a$ excitation energies to the triplet, singlet, and mixed
    symmetry states. The mixed symmetry state has energy $\omega_M=A_a(i^{-1})-I_i$.
    $\Delta$SCF quantities are obtained by the usual
    multiplet sum procedure~\cite{ZRB77} except that a truncated Taylor
    expansion of the xc-functional has been used in the DFT case~\cite{CGG+00}.
    The identification of $I_i$ and $A_a$ in the TD-DFT case is based upon
    OEP theory~\cite{C08}.)
  }
  \label{tab:TOTEM} 
  \begin{center}
  \begin{tabular}{ll}
  \hline \hline
  \multicolumn{1}{c}{$\Delta$SCF HF} & \multicolumn{1}{c}{CIS (TDA-TD-HF)}\\
  $I_i=\epsilon_i$ & $I_i=\epsilon_i$\\
  $A_a=\epsilon_a$ & $A_a=\epsilon_a$\\
  $A_a(i^{-1})=A_a-(aa\vert f_H \vert ii ) +(ai\vert f_H \vert ia)$ &
     $A_a(i^{-1})=A_a-(aa\vert f_H \vert ii ) +(ai\vert f_H \vert ia)$ \\
  $\omega_M = \epsilon_a-\epsilon_i-(aa\vert f_H \vert ii ) +(ai\vert f_H \vert ia)$ &
    $\omega_M = \epsilon_a-\epsilon_i-(aa\vert f_H \vert ii ) +(ai\vert f_H \vert ia)$ \\
  $\omega_T=\omega_M -(ia\vert f_H \vert ai)$  &
    $\omega_T=\omega_M -(ia\vert f_H \vert ai)$ \\
  $\omega_S=\omega_M +(ia\vert f_H \vert ai)$  &
     $\omega_S=\omega_M +(ia\vert f_H \vert ai)$ \\
  \multicolumn{1}{c}{linearized $\Delta$SCF DFT} & \multicolumn{1}{c}{TDA-TD-DFT} \\
  $I_i = \epsilon_i - \frac{1}{2} (ii \vert f_H + f_{xc}^{\uparrow,\uparrow} \vert ii )$ &
     $I_i = \epsilon_i$\\
  $A_a = \epsilon_a + \frac{1}{2} (aa \vert f_H + f_{xc}^{\uparrow,\uparrow} \vert aa )$ &
     $A_a = \epsilon_a+(aa|f_H|ii) +(ai|f_{xc}^{\uparrow,\uparrow}|ia)$\\
  $A_a(i^{-1}) = A_a - (aa \vert f_H + f_{xc}^{\uparrow,\uparrow} \vert ii )$ &
     $A_a(i^{-1}) = A_a -(aa\vert f_H \vert ii ) +(ai\vert f_H \vert ia)$\\
  $\omega_M =
   \epsilon_a-\epsilon_i
         + \frac{1}{2} (aa-ii\vert f_H + f_{xc}^{\uparrow,\uparrow} \vert aa-ii)$ &
   $\omega_M = \epsilon_a-\epsilon_i+(ai\vert f_H + f_{xc}^{\uparrow,\uparrow}\vert ia)$\\
  $\omega_T=\omega_M + (aa \vert f_{xc}^{\uparrow,\uparrow}-f_{xc}^{\uparrow,\downarrow} \vert ii)$ &
    $\omega_T = \omega_M  -(ia\vert f_H + f_{xc}^{\uparrow,\downarrow} \vert ai)$ \\
  $\omega_S=\omega_M - (aa \vert f_{xc}^{\uparrow,\uparrow}-f_{xc}^{\uparrow,\downarrow} \vert ii)$ &
    $\omega_S = \omega_M +(ia\vert f_H + f_{xc}^{\uparrow,\downarrow} \vert ai)$ \\
  \hline \hline
  \end{tabular}
  \end{center}
\end{table}

\marginpar{\hrule $\,$ \\ {\color{blue} STEX}: static exchange\\ \hrule}
Unfortunately the CIS formulae in {\bf Table~\ref{tab:TOTEM}} turn out to be rather bad approximations 
for the singlet and triplet
excitation energies because, as Koopmans' theorem tells us, HF orbital energies are preprepared to describe
ionization and electron attachement rather than neutral excitations.  It would be better to begin with
a zero-order picture where both the initial and target orbitals ``see'' $N-1$ electrons.  This is the idea
behind the static exchange method (STEX) \cite{ACVP97} where the excitations of a neutral molecule that begin
with MO $i$ are carried out by first performing a calculation on the cation with one electron removed from
the MO $i$ (denoted by the abbreviated configuration, $i^{-1}$).  The associated unoccupied cation MO creation 
operators, $a^\dagger(i^{-1})$, are then used together with the 
occupied neutral MO destruction operators, $i$, to carry out a variational calculation with a wave function of the
form $\Psi = \sum_a a^\dagger(i^{-1}) i \Phi_{HF}$.  
Applying this idea to the TOTEM gives the STEX formulae,
\begin{eqnarray}
  \omega_M & = & A_a(i^{-1}) - I_i \nonumber \\
  \omega_S & = & \omega_M +  (ia \vert f_H \vert ai) \nonumber \\
  \omega_T & = & \omega_M  - (ia \vert f_H \vert ai) \, ,
  \label{eq:wft.24}
\end{eqnarray}
where the singlet, $\omega_S$, and triplet, $\omega_T$, excitation energies have been expressed in terms
of an excitation to a fictitious single-determinantal state of mixed symmetry whose excitation
energy, $\omega_M$, is equal to the difference between minus the electron affinity of the $i^{-1}$ cation,
$A_a(i^{-1})$ and minus the ionization potential of the MO $i$, $I_i$.
Specific CIS values are given in {\bf Table~\ref{tab:TOTEM}}.

It is interesting to ask what happens for a charge-transfer excitation between two neutral molecules
located at a distance $R$ apart, where $R$ is assumed to be large.  
According to Eq.~(\ref{eq:wft.24}), we have $\omega_T \approx \omega_M \approx \omega_S $, when the
differential overlap $\psi_i({\bf r}) \psi_a({\bf r})$ becomes negligeable.
It is easy to see from Koopmans' theorem and the CIS results in {\bf Table~\ref{tab:TOTEM}}, that
$\omega_M \approx A_a - I_i - 1/R$ as would be expected on physical grounds.  That is, the excitation
energy required to tranfer an electron from MO $i$ to MO $a$ over a distance $R$ is just the energy 
needed to ionize the electron from MO $i$ plus the energy recuperated
from adding it into MO $a$ and from the electrostatic attraction of the two ions.

\subsection{\color{red} Stability Analysis and Bond Breaking}
\label{subsec:stability}

\marginpar{\hrule $\,$ \\ {\color{blue} SODS}: same-orbitals-for-different-spins\\
{\color{blue} DODS}: different-orbitals-for-different-spins\\ \hrule}
Most HF calculations are really restricted HF calculations with the same-orbitals-for-different-spins (SODS, as
opposed to DODS, which stands for different-orbitals-for-different-spins.)  
In addition, they assume real MOs, each of which belongs to an irreducible representation of the
molecular point group.  Unrestricted HF  consists of dropping the restriction of same orbitals for 
different spin, but still typically uses real MOs.  As unrestricted HF calculations are more general than restricted HF
calculations, cases arise where unrestricted HF calculations give a lower energy than restricted HF 
calculations by allowing
spatial symmetry breaking.  Since the iterations in the HF SCF calculation typically preserve symmetry
for closed-shell molecules, finding unrestricted HF calculations with lower energy than restricted HF 
calculations typically
requires explicit symmetry breaking to be introduced during the iterations.  On the other hand, such
symmetry breaking can, for example, be a simple way to dissociate H$_2$ into neutral atoms, 
$\mbox{H$\uparrow$}+\mbox{H$\downarrow$}$, which is often a good approximation to the correct 
dissociation into a structure, $\left[ \mbox{H$\uparrow$}+\mbox{H$\downarrow$} \leftrightarrow 
\mbox{H$\downarrow$}+\mbox{H$\uparrow$} \right]$, with equal probability for finding each spin
on each atom.  Generalized HF removes all restrictions on the HF orbitals. 
It turns out that there is an intimate relationship between restricted HF stability
analysis and LR-TD-HF.  

Symmetry breaking will occur when an orbital unitary transformation, 
$\psi_r^\lambda({\bf r}) = e^{\lambda \hat{U}^*} \psi_r({\bf r}) = e^{i\lambda(\hat{R}+i\hat{I})} \psi_r({\bf r})$ 
leads to energy lowering,
\begin{eqnarray}
  E_\lambda  & = & E_0 + \frac{\lambda^2}{2} 
  \left( \begin{array}{cc} \vec{U}^{\dagger *} & \vec{U}^\dagger \end{array} \right)
  \left[ \begin{array}{cc} {\bf A} & {\bf B} \\ {\bf B} & {\bf A} \end{array} \right] 
  \left( \begin{array}{l} \vec{U} \\ \vec{U}^* \end{array} \right) + {\cal O}^{(3)}(\lambda) \nonumber \\
  & = & E_0 + \lambda^2 \left[ \vec{R}^\dagger \left({\bf A}-{\bf B}\right) \vec{R} 
  + \vec{I}^\dagger \left({\bf A}+{\bf B}\right) \vec{I} \right] + {\cal O}^{(3)}(\lambda) \, ,
  \label{eq:wft.25}
\end{eqnarray}
where the ${\bf A}$ and ${\bf B}$ matrices are the same as previously defined in Eq.~(\ref{eq:wft.11})
except that they are real because we are beginning with a restricted HF solution with real orbitals.
As Eq.~(\ref{eq:wft.10}) can always be put into the form $\left( {\bf A}+{\bf B}\right) \left( {\bf A}-{\bf B} \right)
\vec{Z} = \omega \vec{Z}$, where $\vec{Z}=\vec{X}-\vec{Y}$,  it is immediately evident that an imaginary 
LR-TD-HF excitation energy means that one or both of the matrices $ {\bf A}+{\bf B}$ and ${\bf A}-{\bf B}$ have negative 
eigenvalues and thus that the restricted HF solution is variationally unstable because there is some
choice of the vectors $\vec{R}$ and $\vec{I}$ which will further lower the restricted HF energy.
The most common of these is the so-called ``triplet instability'' where the imaginary energy
is associated with a triplet excitation energy.  However other types of instabilities are possible.
For a more complete analysis see 
for example Ref.~\cite{F81}.

Note that imaginary excitation energies can never occur when the LR-TD-HF equation is solved in the TDA.
In fact, the variational nature of CIS does much to prevent the unphysical plunges of excitation energies
which are often seen in LR-TD-HF as, say, singlet near instabilities associated with triplet instabilities.
Thus the TDA is a partial solution to the LR-TD-HF instability problem.

Physically LR-TD-HF is giving nonsensical excitation energies because it is based upon the response of
an unphysical ground state wave function.  In fact, while the HF approximation is often a good starting 
point for closed-shell molecules at their ground-state equilibrium geometry --- i.e., it may be reasonable
around the minima on the ground-state curve in {\bf Fig.~\ref{fig:PES}} --- the same approximation is 
notorious for failing at transition states where bonds are being made or broken.  In this case, the 
minimum acceptable description is a wavefunction which is a linear combination of the reactant, $\Phi_R$, 
and product, $\Phi_P$, trial wave functions, $\Phi = \Phi_R C_R + \Phi_R C_R$.  Typically this 
involves some sort of configuration mixing between the ground-state determinant and a two-particle/two-hole
(doubly) excited state.  Though there is no transition state on the ground state curve, the dissociation
of the H($1s$)-H($1s$) $\sigma$-bond is a textbook case.  The minimum acceptable correctly dissociating
description of the ground state wave function, satisfying all the spatial and spin symmetry rules, is
a trial function of the form,
$\Psi = \vert \sigma_g \bar{\sigma}_g \vert C_1 + \vert \sigma_u \bar{\sigma}_u \vert C_2 $,
where $\sigma_g$ is the familiar bonding MO and $\sigma_u$ is the familiar antibonding MO.
The coefficients $C_1 = 1/\sqrt{2}=-C_2$ give the proper neutral,
$\mbox{H$_2$} \rightarrow \left[ \mbox{H$\uparrow$} + \mbox{H$\downarrow$} \leftrightarrow
\mbox{H$\downarrow$} + \mbox{H$\uparrow$} \right]$, dissociation limit of the ground
$X \, ^1\Sigma_g$ state.  The combination $C_1 = 1/\sqrt{2}=C_2$ gives the proper ionic dissociation limit,
$\mbox{H$_2$} \rightarrow \left[ \mbox{H:$^-$} + \mbox{H$^+$}
\leftrightarrow \mbox{H$^+$} + \mbox{H:$^-$} \right]$, of the ``two-particle/two-hole'' (``2p2h'') $1 \, ^1\Sigma_g$ state.
Though the $1 \, ^1\Sigma_g$ state
is dominated in this simple model by the 2p2h configuration at the ground-state equilibrium geometry,
the fact that neither $C_1$ nor $C_2$ are zero in the dissociation limit indicates that configuration
mixing of the ground and 2p2h excited-state configurations is occuring.  In fact the simplest
wave-function model for breaking of any homoleptic single bond involves mixing of the ground-state
configuration with a 2p2h configuration.

\section{{\color{red} CLASSIC DENSITY-FUNCTIONAL THEORY}}
\label{sec:dft}

It is difficult (possibly impossible) to speak of TD-DFT without first saying a word about ordinary DFT.
This is the classic DFT of the ground stationary state.  It should also be the static limit of TD-DFT.
Since our focus in this chapter is not on DFT, which is already fairly well known, but on TD-DFT, we shall 
try to keep this section short.  Readers interested in more information about DFT may choose to consult 
any one of a number of books on the subject such as, for example, that of Dreizler and Gross \cite{DG90}.

\marginpar{\hrule $\,$ \\ {\color{blue} DFA}: Density-functional approximation. \\ \hrule}
Before going further, it is important to make the distinction between DFT (density-functional {\it theory})
and density-functional {\it approximations} (DFAs).  DFT is a formalism which provides existence theorems
showing that there is something well defined to be approximated.  However from a practical point of view,
DFT is pretty useless unless practical DFAs can be found which are simpler to use than the wave-function theory
of corresponding accuracy.  The search for pragmatic approximations, ideally guided by rigorous formal 
DFT but occasionally borrowing from wave-function theory, pervades all of density-functional culture.

\subsection{\color{red} Formalism}
\label{sec:dft_formalism}

Hohenberg and Kohn set down the basic formalism of modern DFT in two theorems \cite{HK64}.  The {\it first
Hohenberg-Kohn theorem} states that the ground state charge density of a system of interacting electrons determines
the external potential up to an arbitrary additive constant.  Since integrating over the charge density
gives the number of electrons, this implies that the ground-state charge density determines the Hamiltonian
up to an additive constant (i.e., the arbitrary zero of the energy) and hence determines pretty much
everything about the ground- and excited-states of the system.  

In fact, the ground state density must even determine the {\it dynamic} linear response of the ground stationary state,
\begin{equation}
  \chi({\bf r}_1,{\bf r}_2;\omega) = \int_{-\infty}^{+\infty} e^{+i\omega (t_1-t_2)} 
             \frac{\delta \rho({\bf r}_1,t_1)}{\delta v_{appl}({\bf r}_2,t_2)} \, d(t_1-t_2) \, ,
  \label{eq:dft.1}
\end{equation}
since this can be expressed in terms of the ground- and excited-state wave functions and excitation energies,
\begin{equation}      
  \chi({\bf r}_1,{\bf r}_2;\omega) =
   \sum_{I\neq 0} \left( \frac{\langle 0 \vert \hat{\rho}({\bf r}_1) \vert I \rangle 
    \langle I \vert \hat{\rho}({\bf r}_2) \vert 0 \rangle}{\omega-\omega_I+i\eta} 
  - \frac{\langle 0 \vert \hat{\rho}({\bf r}_2) \vert I \rangle
    \langle I \vert \hat{\rho}({\bf r}_1) \vert 0 \rangle}{\omega+\omega_I+i\eta}  \right) \, ,
  \label{eq:dft.2}
\end{equation}
where $\hat{\rho}$ is the density operator and $\eta=0^+$ is an infinitesimal, needed to guarantee
causality: $\delta \rho({\bf r}_1,t_1)/\delta v_{appl}({\bf r}_2,t_2) = 0$ if $t_2 > t_1$.

Counterbalancing this marvelous theorem are the representability problems.  That is, given
a candidate ground-state charge density which integrates to $N$ electrons: (i) How can we be 
sure that it comes from an $N$-electron wave function (pure-state $N$ representability) 
or at least from an $N$-electron ensemble density matrix (ensemble $N$ representability)? 
(ii) And how can we be sure that it is the ground-state density for some external potential
($v$ representabilty)?  It turns out, with few exceptions, that most reasonable candidate
charge densities are pure-state $N$ representable, but determining the $v$-representability
of a candidate charge density is far from obvious.

There is also the problem of how to actually calculate something nontrivial such as the ground-state
energy from a candidate charge density.  The {\it second Hohenberg-Kohn theorem} solves this problem
by using the variational principle to show that, $E_0 \leq F[\rho] + \int v_{ext}({\bf r}) \rho({\bf r}) \, d{\bf r}$,
where the functional $F[\rho]$ is universal in the sense that it is independent of the external 
potential.  Equality is achieved for the ground-state charge density.  In this case, we can get
around the $v$-representability problem using the Levy-Lieb constrained search formalism,
$F[\rho] = \inf_{\Psi \rightarrow \rho} \langle \Psi \vert \hat{T} + \hat{V}_{ee} \vert \Psi \rangle$, which
says to search over all wave functions (or ensemble density matrices) which produce the trial density.
In this sense, the Hohenberg-Kohn functional is {\it not} unknown, but it is too complicated to be
practical and so approximations must be sought.

Such approximations fall under the category of orbital-free theories, of which Thomas-Fermi theory
is the classic historical example.  They tend not to be as accurate as approximations based upon the
Kohn-Sham (re)formulation of DFT. 
The main problem in making a DFA for an orbital-free theory is in finding an approximation for the 
kinetic energy.  The Kohn-Sham formulation of DFT gets around this problem introducing a fictitious
system of {\it noninteracting} electrons with the same charge density as the real system of $N$
interacting electrons.  The total energy expression is exactly the same as that in the HF approximation 
[Eq.~(\ref{eq:wft.5})] except that the HF exchange energy, $E_x[\gamma]$, is now replaced with
the exchange-correlation (xc) functional, $E_{xc}[\rho]$, and the total electronic energy is 
exact when the xc-functional is exact (in contrast to HF which is always an approximation).
Note that the xc-functional, $E_{xc}[\rho]= F[\rho] - E_H[\rho] 
- \inf_{\Phi \rightarrow \rho} \langle \Phi \vert \hat{T} \vert \Phi \rangle$ where $\Phi$ is 
a single-determinantal wave function,
contains not only the exchange and correlation energies familiar
from wave function theory but also the difference of the kinetic energies of the interacting 
and interacting systems.  Also, before going any further, let us follow modern practice and 
generalize the original Kohn-Sham DFT to spin DFT so that $E_{xc}=E_{xc}[\rho_\alpha,\rho_\beta]$.
Minimizing the total energy with respect to the orthonormal orbitals 
of the noninteracting system then leads to the Kohn-Sham equation which differs only from the 
corresponding HF orbital equation (\ref{eq:wft.3}) by the replacement of the HF exchange self-energy 
[Eq.~(\ref{eq:wft.4})] by the xc-potential,
\begin{equation}
  v_{xc}^\sigma[\rho_\alpha,\rho_\beta]({\bf r}) = \frac{\delta E_{xc}[\rho_\alpha,\rho_\beta]}{\delta \rho_\sigma({\bf r})} \, .
  \label{eq:dft.3}
\end{equation}

\marginpar{\hrule $\,$ \\ {\color{blue} PNDD}: particle number derivative discontinuity \\ 
{\color{blue} OEP}: optimized effective potential \\
{\color{blue} EXX}: exact exchange \\
{\color{blue} KLI}: Krieger-Li-Iafrate \\
\hrule}
It is now well established that there is a particle number derivative
discontinuity (PNDD) in the exact xc-potential, whereby adding a fraction of an electron into a higher-energy
orbital leads to a drastic change in the asymptotic behavior of the xc-potential and a nearly rigid shift
of the xc-potential in the ``bulk region'' where the density is large.  
This may be explicitly verified within the optimized effective potential (OEP) model which is most
easily understood in its classic exchange-only version where it is often called exact exchange (EXX).
The EXX model finds the potential whose orbitals minimize the HF energy expression.  Alternatively the OEP
potential is that for which the linear response of the density is zero when replacing the HF exchange self-energy
with a local potential.  Thus the OEP provides an excellent approximation to the true Kohn-Sham x-potential,
complete with PNDD.  This is easiest to see within the Krieger-Li-Iafrate (KLI) approximation to the EXX
potential for which we will now give a heuristic derivation: 
Let us assume that the HF and KS orbitals are about the same {\it on average}.  If they really were identical,
then we could immediately write that,
$\left( \hat{\Sigma}_x - v_x({\bf r}) \right) \psi_i({\bf r}) =  \left( \epsilon_i^{HF} - \epsilon_i^{KS} \right) \psi_i({\bf r})$.
However this is only supposed to be true on average so we should multiply by $n_i \psi_i^*({\bf r})$ and sum over the index $i$.
After rearrangement, this leads to,
\begin{equation}
  v_x({\bf r}) = \frac{\sum_i n_i \psi_i^*({\bf r}) \hat{\Sigma}_x \psi_i({\bf r})}{\rho({\bf r}) }
              + \frac{\sum_i n_i  \left( \epsilon_i^{HF} - \epsilon_i^{KS} \right) \vert \psi_i({\bf r}) \vert^2}{\rho({\bf r})}
  \, .
  \label{eq:tddft.16}
\end{equation}
The first term on the right-hand-side is Slater's historic definition of the local exchange potential.  The second term is
the PNDD correction.  It is there because the asymptotic behavior of the charge density depends upon the value of HOMO energy.
\marginpar{\hrule $\,$ \\ 
{\color{blue} NVR}: noninteracting $v$-representability. \\ \hrule}
This second term introduces a rigid shift to compensate for the change in the asymptotic behavior every time a new HOMO is
occupied.  Such discontinuous behavior is very difficult to incorporate into practical DFAs other than by an OEP-like approach.

With the Kohn-Sham formulation, DFT also gains a new representability problem, namely the problem
of noninteracting $v$-representability (NVR).  This is the requirement that there be a noninteracting
system {\it integer occupation numbers} whose ground state has the same density as that of the 
interacting system.  Violation of NVR would show up as a ``hole below the Fermi level'' meaning
\marginpar{\hrule $\,$ \\ 
{\color{blue} HOMO}: Highest occupied molecular orbital. \\
{\color{blue} LUMO}: Lowest unoccupied molecular orbital. \\ \hrule}
that the lowest unoccupied molecular orbital (LUMO) is lower in energy than the highest occupied molecular
orbital (HOMO) in violation of the assumed Fermi statistics for the ground state of the noninteracting
system.  Now suppose every system were NVR. Then we should always have an exact spin-restricted
(same-orbitals-for-different-spin) solution for dissociating molecules with a singlet ground state
belonging to the totally-symmetric representation of the molecular symmetry group.  That is, symmetry
breaking should never occur for the exact functional.  This is because the
exact ground state spin $\alpha$ and spin $\beta$ densities are identical, implying that 
$v_{xc}^\alpha({\bf r})=v_{xc}^\beta({\bf r})$, and hence that the orbitals can always be chosen
identical for the different spins \cite{CJI+07}.  It now seems likely that not every system is NVR 
(see Ref.~\cite{TTR+08} and references therein).  This leads naturally to fractional occupation numbers
as is easily seen in the TOTEM.  Consider a spin-wave instability, where the Kohn-Sham determinant
takes the form, $\Psi_\lambda = \vert \sqrt{1-\lambda^2} \psi_i + i \lambda \psi_a , 
\sqrt{1-\lambda^2} \bar{\psi}_i + i \lambda \bar{\psi}_a \vert$, where the absence (presence) of an
overbar indicates spin $\alpha$ ($\beta$). The total energy may then be expanded as $E_\lambda =
E_0 + 2 \lambda^2 (\epsilon_a - \epsilon_i) + {\cal O}(\lambda^3)$.  The spin-wave instability
will lower the energy whenever the LUMO energy, $\epsilon_i$, is lower than the HOMO energy, $\epsilon_a$.
(See Ref.~\cite{O72} for a similar example within the HF model.)
However it is then easy to demonstrate that the orbitals have fractional occupation number by looking at 
the corresponding 1-RDM operators,
\begin{eqnarray}
  \hat{\gamma}_\alpha & = & (1-\lambda^2) \vert i \rangle \langle i \vert + i \lambda \sqrt{1-\lambda^2} \left( \vert a \rangle \langle i \vert
                             - \vert i \rangle \langle a \vert \right) + \lambda^2 \vert a \rangle \langle a \vert
  \nonumber \\
  \hat{\gamma}_\beta & = & (1-\lambda^2) \vert i \rangle \langle i \vert - i \lambda \sqrt{1-\lambda^2} \left( \vert a \rangle \langle i \vert
                             - \vert i \rangle \langle a \vert \right) + \lambda^2 \vert a \rangle \langle a \vert
   \nonumber \\
  \hat{\gamma} & = & \hat{\gamma}_\alpha + \hat{\gamma}_\beta = 2 (1-\lambda^2) \vert i \rangle \langle i \vert
                    + 2 \lambda^2 \vert a \rangle \langle a \vert \, .
  \label{eq:dft.4}
\end{eqnarray}

These observations are consistent with the traditional assumption that an ensemble formalism with fractional 
occupation number is needed in cases where NVR fails.   An important theorem in this context is that all 
fractionally-occupied orbitals must be degenerate with lower energy orbitals being fully occupied and higher-energy
orbitals being completely vacant (see e.g., Ref.~\cite{DG90}, pp.\ 55-56).  Surprisingly there is some indication
that the noninteracting system may become degenerate in such a theory for a range of molecular configurations where the
interacting system is nondegenerate \cite{SGB99}.  One is then possibly faced with the problem of minimizing the energy 
with respect to 
unitary transformations within 
the space spanned by the degenerate occupied orbitals with different occupation numbers. 

\subsection{\color{red} Approximations}
\label{sec:dft_approximation}

\begin{figure}
  \begin{center}
    \includegraphics[angle=0,width=1.0\textwidth]{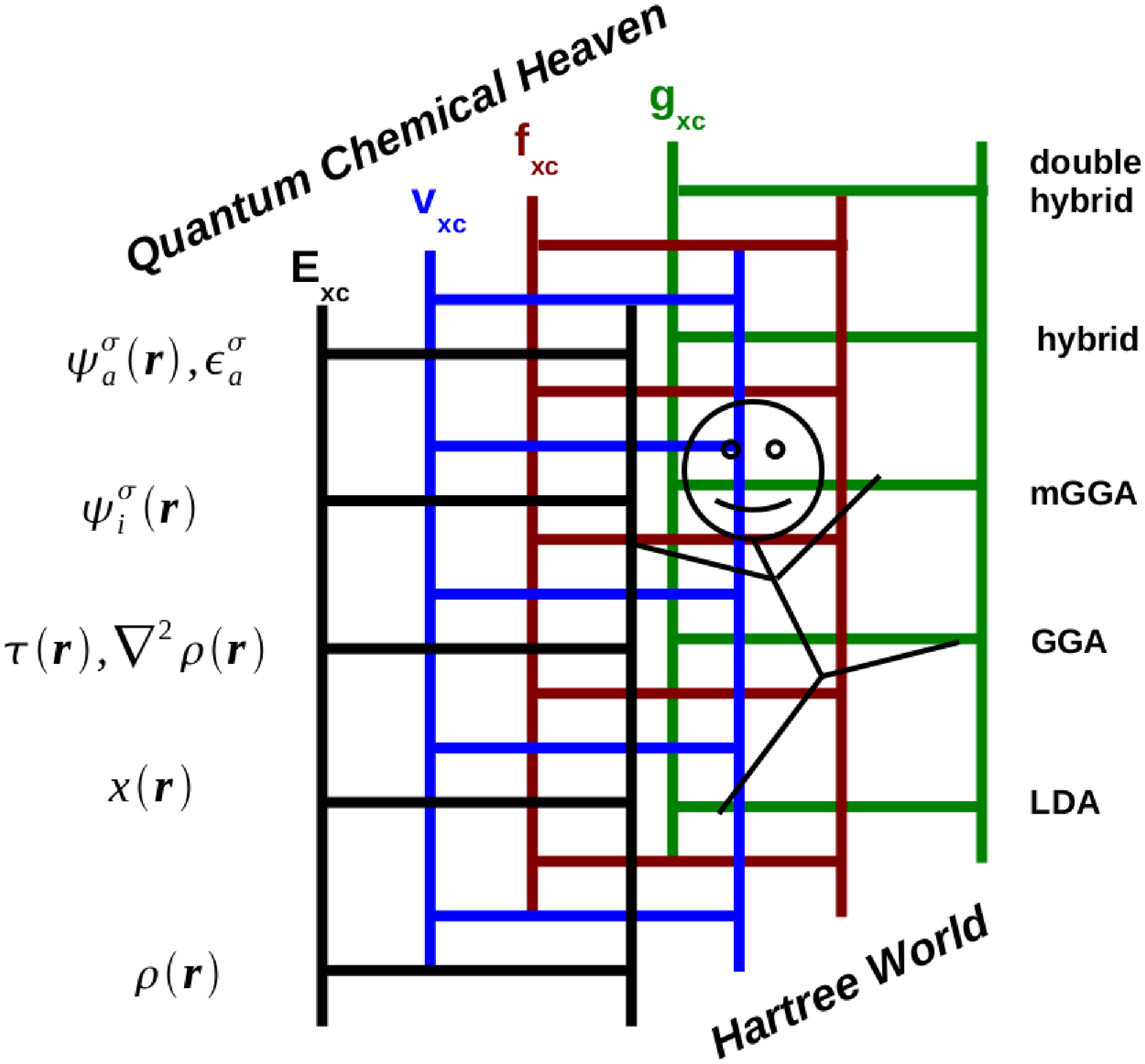}
  \end{center}
  \caption{\label{fig:junglegym}
      The front (black) ladder of Jacob's jungle gym \cite{C02} is Jacob's ladder for $E_{xc}$ \cite{PS01}.  
      Spin indices have been dropped 
      for simplicity, however present practice is to always use spin-density functional theory where the density 
      is a two-component object, $(\rho_\alpha,\rho_\beta)$. {\color{blue} LDA}, local density approximation, 
      involves only the
      density, $\rho({\bf r})$, at each point; {\color{blue} GGA}, generalized gradient approximation, 
      involves both $\rho({\bf r})$
      and the reduced gradient, $x({\bf r})=\vert \vec{\nabla} \rho({\bf r}) \vert/ \rho^{4/3}({\bf r})$; 
      {\color{blue} mGGA}, meta generalized gradient approximation, involves $\rho({\bf r})$, $x({\bf r})$, and the local
      kinetic energy, $\tau({\bf r}) = \sum_p n_{p} \psi_{p}({\bf r}) \nabla^2 \psi_{p}({\bf r})$, or the 
      Laplacian of the density, $\nabla^2 \rho({\bf r})$; 
      in climbing to the fourth rung --- that of 
      hybrid functionals, exact exchange, and related functionals --- explicit information about occupied orbitals
      is also incorporated into the functional; finally on the fifth and highest rung --- that of double hybrids, 
      functionals based upon the adiabatic connection fluctuation-dissipation theorem, and related functionals ---
      explicit information is added about unoccupied orbitals and their orbital energies.  Effectively at the
      highest level DFT closely approaches conventional many-body theory.
      }
\end{figure}

DFT by itself is 
hardly useful without DFAs.
John Perdew has organized functionals into a Jacob's Ladder 
(Fig.~\ref{fig:junglegym}).  The basic idea is that gradually including more information about the local density, the 
reduced gradient, the local kinetic energy, etc.\ allows users (angels going up the ladder) the added flexibility 
needed to create more accurate functionals.  
\marginpar{\hrule $\,$ \\ {\color{blue} Pure density functional}: One depending only upon the density, excluding any
orbital dependence.\\ \hrule}
Only the first two rungs (or three if the local kinetic-energy is excluded from the third rung) constitute pure
density functionals --- that is, functionals which stick to the original spirit of Hohenberg-Kohn-Sham DFT and
consider only the density as the working variable.  For this reason it might be better to refer to the higher rungs
as generalized Kohn-Sham.  It must be emphasized that, while greater accuracy is not guaranteed, 
there is a general tendency towards more accurate functionals when going to higher rungs of the ladder.  However it 
is important that users also be able to descend the ladder when necessary as calculations necessarily become more 
complex and expensive on the higher rungs of the ladder.

One of us has generalized Jacob's ladder to Jacob's jungle gym \cite{C02} by including additional ladders corresponding
to higher xc-derivatives.  For pure density functionals these would be defined by Eq.~(\ref{eq:dft.3})
and,
\begin{eqnarray}
  f_{xc}^{\sigma_1,\sigma_2}({\bf r}_1, {\bf r}_2) & = & \frac{\delta^2 E_{xc}[\rho_\alpha,\rho_\beta]}
                           {\delta \rho_{\sigma_1}({\bf r}_1) \delta \rho_{\sigma_2}({\bf r}_2)} \nonumber \\
  g_{xc}^{\sigma_1,\sigma_2,\sigma_3}({\bf r}_1,{\bf r}_2,{\bf r}_3) & = & \frac{\delta^3 E_{xc}[\rho_\alpha,\rho_\beta]}
               {\delta \rho_{\sigma_1}({\bf r}_1) \delta \rho_{\sigma_2}({\bf r}_2) \delta \rho_{\sigma_3}({\bf r}_3)}
   \, ,
  \label{eq:dft.5}
\end{eqnarray}
where the xc-kernel, $f_{xc}$, and  xc-hyperkernel, $g_{xc}$, are needed for static response properties.  There
are other analogous quantities at each of the higher rungs.  This type of representation is useful to keep in mind
because derivative quantities are always harder to approximate than the quantities whose derivative is being taken.
It may thus sometimes be appropriate to combine different approximations from different ladders when calculating certain 
types of quantities.

\subsection{\color{red} Scale-up}
\label{sec:dft_scaleup}

Since Kohn-Sham DFT resembles the HF approximation which is the basis of much of the development of (non-DFT)
{\it ab initio} quantum chemistry, it is relatively easy to compare the two approaches and even to incorporate
a Kohn-Sham solver into a wave-function {\it ab initio} approach.  This makes especially good sense if DFAs 
are viewed as a way of extending the accuracy of wave-function {\it ab initio} approaches to larger systems
then would otherwise be possible.  In this sense, the problem of how to scale-up calculations to ever larger 
systems is highly relevant to the success of DFT and DFAs.  However space limitations prevent us from saying 
more than only a few words about scale-up strategies.

It has long been noted that the lower rungs of Jacob's ladder (i.e., the LDA and the GGAs) more closely resemble
the Hartree than the HF approach.  As such, rather than have a formally ${\cal O}^{(4)}(N)$ scaling, the introduction
of density-fitting functions could lead to a formal ${\cal O}^{(3)}(N)$ scaling which could be further improved
upon by integral prescreening.  Beyond a certain system size (and hence once a certain level of computational
power is reached), special algorithms can be used such as those which scale asymptotically as ${\cal O}(N)$.
These algorithms are especially easy to apply in the case of DFT and makes linear-scaling DFT extremely attractive 
for large-scale ``chemistry on the computer'' \cite{HA08}.

A second scale-up strategy is the subsystem approaches where formal DFT \cite{WW93,W08} is extended to describe subsystems of
larger systems --- such as molecules hydrogen bonded to other molecules, molecules in solution,
or even fragments of molecules covalently bonded to the whole --- and corresponding DFAs are developed.  
Not only does such an approach allow the treatment of larger systems by allowing a Kohn-Sham description
of a subsystem to be embedded within a larger surrounding system described, say, by using an orbital-free DFA, but
the subsystem approach is also attractive from an interpretational point of view as it corresponds nicely,
say,  with the idea of functional groups in organic chemistry and active sites in biochemistry.
An intriguing development is the successful implementation of subsystem density-functional theory for the
case that the subsystem is covalently bonded to the larger system \cite{CW04,JV08}.

\marginpar{\hrule $\,$ \\ {\color{blue} DFTB}: Density-functional tight-binding. \\ \hrule}
A final strategy which is worth mentioning is the development of the density-functional tight-binding (DFTB)
model with parameters obtained from Kohn-Sham calculations \cite{FSE+00}.  While loosing some of the rigor of a true Kohn-Sham
calculation, DFTB is a next logical step in a multiscale strategy to larger spatial and temporal scales than can normally
be obtained by true Kohn-Sham calculations.

\section{{\color{red} TIME-DEPENDENT DENSITY-FUNCTIONAL THEORY}}
\label{sec:tddft}

The classic Hohenberg-Kohn-Sham DFT is a theory of the ground stationary state.  However, as mentioned
in the introduction, chemistry is not just a matter of the ground stationary state, but also of UV-Vis 
spectroscopy, NLO, photochemistry, and other applications involving either time-dependent electronic
properties and/or electronic excited states.

\marginpar{\hrule $\,$ \\ {\color{blue} $\Delta$SCF}: Method consisting of calculating the energy difference
of two SCF calculations. \\ \hrule}
In 1999, Singh and Deb reviewed the prospects for treating excited states within a DFT framework \cite{SD99}.
Up to 1995 \cite{C95}, the most popular method was the $\Delta$SCF approach and its MSM variant.  Since
the classic Hohenberg-Kohn-Sham DFT can handle electronic ground states, it is possible to carryout
self-consistent field (SCF) calculations for the ground states of a neutral molecule, its anion, and its cation.
By taking energy differences, one has a rigorous method for calculating the first ionization potential and
electron affinity.  By adding or removing electrons from other orbitals, one has at least an ad hoc way to 
estimate other ionization potentials and electron affinities.  For some of these ionization potentials and 
electron affinities, the method may even be considered to be reasonably rigorous to the extent that we can accept 
that the DFAs developed for the true ground states also should apply to the ground state of each symmetry, 
particularly if a single-determinant is a reasonable first approximation.  Within the same assumption, the 
lowest triplet excitation energy of a closed-shell molecule may also be calculated reasonably rigorously
within the $\Delta$SCF method.  For open-shell singlet excited states, the Ziegler, Rauk, and Baerends \cite{ZRB77}
and Daul \cite{D94} adapted the MSM method as an excited-state DFA.  The result in the TOTEM is given
in {\bf Table~\ref{tab:TOTEM}}  after linearization.
The difference is that minus the
ionization potential, $I_i$, of orbital $i$ and minus the electron affinity, $A_a(i^{-1})$, are given by
expresions which are very different than those given by Koopmans' theorem.
In fact, unlike the case of Hartree-Fock, one would expect Kohn-Sham orbitals
to be preprepared to describe excitations in the sense that for pure DFAs, the occupied and unoccupied
orbitals ``see'' the same potential and hence the same number of electrons.  This is particularly clear
in the expression for the mixed-state excitation energy, $\omega_M$, 
which is just a simple orbital energy difference when the densities of the initial ($i$) and final ($a$) orbitals are
the same.  The additional term is a correction for relaxation of the orbital densities during the excitation process.
There is thus a lot of good physics in the MSM DFA for excited states.  However the major difficulty with this method is
that the form of the excited-state wave function must be estimated, typically based upon symmetry arguments.  It thus
only a first-order estimate in many cases and is also more difficult to automate for excited-state calculations than
is TD-DFT.

\subsection{\color{red} Classic TD-DFT}
\label{sec:tddft_classic}

The question is to what extent the time-evolution of the charge density can be treated
without having to solve the full the time-dependent Schr\"odinger equation,
\begin{equation}
  \hat{H}(t) \Psi(t) = i \frac{\partial}{\partial t} \Psi(t) \, ,
  \label{eq:tddft.4}
\end{equation}
for the $N$-electron wave function, $\Psi(t)=\Psi(1,2,\cdots,N,t)$.  Before discussing this question, it is useful 
to recall that the time-dependent problem is an initial value problem: Given $\Psi_0=\Psi(t_0)$, propagate
the wave function forward in time to obtain $\Psi(t)$.  Thus the wave function at time $t$ is a functional of
the wave function at time $t_0$, $\Psi[\Psi_0](t)$.
 
Since the time-dependent $N$-electron Schr\"odinger equation can rarely be solved exactly, approximate solutions
are often sought.  This is typically done using the Frenkel-Dirac stationary action principle \cite{D30,F34,LEK72},
which states that making the action,
\begin{equation}
  A[\Psi;\Psi_0](t,t_0)
   = \int_{t_0}^{t} \left( \langle \Psi[\Psi_0](t') \vert i \frac{\partial}{\partial t'} \vert \Psi[\Psi_0](t') \rangle 
     -E[\Psi_0](t') \right) \, dt'
  \, ,
  \label{eq:tddft.5}
\end{equation}
stationary with respect to variations of $\Psi[\Psi_0](t')$ on the interval $t' \in (t_0,t)$ and subject to the constraints
$\delta \Psi[\Psi_0](t_0) = \delta \Psi[\Psi_0](t) = 0$, implies that the time-dependent Schr\"odinger
equation is satisfied over the time interval $(t_0,t)$.  Here,
\begin{equation}
  E[\Psi_0](t) = \langle \Psi[\Psi_0](t) \vert \hat{H} \vert \Psi[\Psi_0](t) \rangle  \, ,
  \label{eq:tddft.6}
\end{equation}
is the ``instantaneous energy'' which can in principle also contain information about system dynamics from
previous times.   This emphasizes the main point of this stationary action principle, mainly that 
an {\it approximate} equation of motion may be developed from any {\it approximate} instantaneous energy expression.  
Vignale pointed out a problem with the original Frenkel-Dirac principle, namely that causality (i.e., $\Psi[\Psi_0](t)$ is 
determined by the boundary condition $\Psi_0$) means that we are not in general free to set $\delta \Psi[\Psi_0](t)=0$.  
When this is taken into account the new Frenkel-Dirac-Vignale stationary action principle \cite{V08} is that,
\begin{equation}
  \delta A[\Psi;\Psi_0](t,t_0) = i \langle  \Psi[\Psi_0](t,t_0) \vert \delta \Psi[\Psi_0](t,t_0) \rangle \, , 
  \label{eq:tddft.7}
\end{equation}
subject only to the obvious constraint that $\delta \Psi[\Psi_0](t_0)=0$ implies that the time-dependent Schr\"odinger equation 
is satisfied over the time interval $(t_0,t)$.  Note that the right hand side of Eq.~(\ref{eq:tddft.7}) set to zero in the original
Frenkel-Dirac formulation.  Fortunately the two stationary action principles often give the same result \cite{V08}, 
which is good since the Vignale correction is rarely invoked.  Note also that Lagrange multipliers may easily be added to 
apply the constraints frequently  needed when deriving equations within a given model.

Now let us return to the question of the extent to which the time-dependent wave function may be replaced by the time-dependent 
charge density.  It is natural to replace the instantaneous energy in the Frenkel-Dirac action with the Kohn-Sham energy
and the wave function with the Kohn-Sham determinant.  The Frenkel-Dirac stationarity condition then leads to the time-dependent
Kohn-Sham equation,
\begin{equation}
  \left[ -\frac{1}{2} \nabla^2 + v_{ext}(t) + v_H[\rho(t)] + v_{xc}[\rho_\alpha(t),\rho_\beta(t)] \right] \psi_i(t)
  = i \frac{\partial \psi_i(t)}{\partial t} \, ,
  \label{eq:tddft.8}
\end{equation}
and this equation was used as early as 1980 \cite{ZS80} to calculate photoabsorption cross-sections of rare gas atoms. 
However the formal justification did not come until later.

Formal TD-DFT is usually traced back to the classic paper of Runge and Gross \cite{RG84} which tried to firm up earlier
work on the same subject. 
Thus Runge-Gross TD-DFT is two decades younger than its stationary ground-state older sibling, the Hohenberg-Kohn-Sham
theory.  The Runge-Gross paper actually contains four theorems.  The first is {\it
the} Runge-Gross theorem which states that the time-dependent charge density, $\rho({\bf r},t)$, together with the 
initial wave function, $\Psi_0$, determines the external potential up to an additive function of time.  This implies
that the wave function is determined up to an arbitrary phase factor, $\Psi(t) = e^{-i\phi(t)} \Psi[\rho,\Psi]$.  The
proof proceeds in two steps.  In the first step, it is shown that any Taylor-expandable time-dependent external potential 
is determined by $\Psi_0$ and by the current density, ${\bf j}({\bf r},t)$.  In the second step, it is to show that
$\rho({\bf r},t)$ is determined by ${\bf j}({\bf r},t)$.  A weakness in the second step of the proof was noted by
Xu and Dhara \cite{XD85} but can be excluded for physically-reasonable external potentials which may be viewed as
being made up of point charges \cite{DG87,GK90}.  {\it Although the TD-DFT community is very self-critical, the proof
of this theorem has thus far withstood the test of time.}  The second theorem in the Runge-Gross paper proposes
a hydrodynamical formulation of TD-DFT.  The third theorem proposes the Frenkel-Dirac action can be made the basis
for a variational theorem somewhat analogous to the second Hohenberg-Kohn theorem.  Unfortunately this theorem is
wrong as stated because (as is always the case with the Frenkel-Dirac principle) it does not handle causality correctly.
Many subsequent reformulations of the action have been proposed including one by Rajagopal involving the Berry
phase \cite{R96}, formulations involving the Keldysh action \cite{L98,L01}, and Liouville space pathways \cite{M05b}.
However the use of the Frenkel-Dirac-Vignale action \cite{V08} seems to be by far the simplest solution.  The fourth
theorem in the Runge-Gross paper proposes a Kohn-Sham formulation of TD-DFT.  This is actually a very significant
statement because it clarifies the nature of the time-dependent xc-potential.  Whether written as the derivative
of an xc-action,
\begin{equation}
  v_{xc}^\sigma[\rho_\alpha,\rho_\beta;\Psi_0,\Phi_0]({\bf r},t) = \frac{\delta A_{xc}[\rho_\alpha,\rho_\beta;\Psi_0,\Phi_0]}
               {\delta \rho_\sigma({\bf r},t)} \, ,
  \label{eq:tddft.9}
\end{equation}
or simply postulated to exist, the xc-potential in time-dependent Kohn-Sham theory has a formal dependence on the 
initial wave functions of the real interacting, $\Psi_0$, and noninteracting Kohn-Sham, $\Phi_0$, systems of $N$-electrons.
This may be eliminated using the first Hohenberg-Kohn theorem if the initial state of the system is the ground stationary
state.  Even so, we do not recover the xc-potential of Eq.~(\ref{eq:tddft.8}),
where time is merely considered to be a parameter and the derivative is taken at fixed time.  This is the important
AA of TD-DFT which assumes that the xc-potential reacts instantaneously and without memory
to any temporal change in the charge density.  The adiabatic approximation is a local approximation in time.  The
exact TD-DFT xc-potential should have memory in the sense of depending upon the past history of the charge density
and is thus a functional of a function of not just $({\bf r},\sigma)$ but also of $t$.  The adiabatic approximation
may be taken as the definition of conventional TD-DFT and in practical terms predated the Runge-Gross paper.  {\it The 
gauntlet thrown down by Runge and Gross to challenge us is to find how to include memory effects in a time-dependent 
xc-functional.}

\marginpar{\hrule $\,$ \\ {\color{blue} CDFT}: current density-functional theory \\ \hrule}
The Runge-Gross formulation of TD-DFT was limited only to applied electric fields.  This is sufficient for many 
problems because magnetic field effects are often significantly smaller than electric field effects in experimental
studies of molecular properties, but it would be better to have a gauge-invariant theory treating electric and
magnetic fields on the same footing.  Ghosh and Dhara have extended TD-DFT to handle time-dependent magnetic fields
by introducing a current-density dependence and an exchange-correlation vector potential \cite{GD88}.
Since that time, TD current density-functional theory (CDFT) has developed considerably (see, e.g., Ref.~\cite{F06}
for one recent paper.)
Alternatively it is possible to formulate and carryout calculations
within the framework of 4-component relativistic TD-DFT \cite{SJ03}.

\subsection{\color{red} Applications}
\label{sec:tddft_applications}

In the quarter century since the Runge-Gross paper \cite{RG84} and the introduction of TD-DFT into quantum chemistry over
a decade ago \cite{C95}, there have been a very large number of interesting applications of this theory.  We will only
try to highlight a few here that we find to be especially interesting.

\subsubsection{\color{red} Absorption Spectra}

The major application of TD-DFT is for the calculation of electronic absorption spectra using linear response theory.
Earlier (Sec.~\ref{sec:dft_formalism}) we argued that the first Hohenberg-Kohn theorem guarantees that the dynamic 
linear response of a system of $N$-electrons is determined by the ground-state charge density, but we had no practical 
way to construct this response.  TD-DFT provides just such a constructive procedure through the dynamic linear response 
of the charge density.  The calculation consists of propagating the Kohn-Sham orbitals in the presence of a small dynamic 
perturbation, $v_{appl}({\bf r},t) = {\bf {\cal E}}(t) \cdot {\bf r}$, and calculating the induced dipole moment, 
${\bf \mu}_{ind}(t)$, which is the difference between the time-dependent and permanent dipole moments of the system.  
The dynamic polarizability, ${\bf \alpha}$ is defined by the linear term, 
${\bf \mu}_{ind}(t) = \int {\bf \alpha}(t-t') {\bf {\cal E}}(t)$.  Applying the Fourier transform convolution theorem 
allows us finally calculate the dynamic polarizability as, ${\bf \alpha}(\omega) = {\bf \mu}(\omega)/{\bf {\cal E}}(\omega)$.  
\marginpar{\hrule $\,$ \\ {\color{blue} SOS}: Sum-over-states. \\ \hrule}
Since the dynamic polarizability has the sum-over-states (SOS) form,
\begin{equation}
  \alpha(\omega) = \sum_{I\neq 0} \frac{f_I}{\omega_I^2-\omega^2} \, ,
  \label{eq:tddft.11}
\end{equation}
where $\omega_I$ and $f_I$ are respectively vertical excitation energies and the corresponding oscillator
strength, the Laurentzian broadened stick spectrum can finally be obtained as,
\begin{equation}
  {\cal S}(\omega) = \frac{2\omega}{\pi} \Im m \, \alpha(\omega+i\eta) \, .
  \label{eq:tddft.12}
\end{equation}
Such a procedure has been implemented, for example, in the {\bf Octopus} code \cite{CMA+06}, and has the 
advantage of being able to go to calculate the absorption spectrum of a very large molecule molecule over 
a wide range of energies, albeit with only moderate spectral resolution.

Alternatively one of us \cite{C95} showed how LR-TD-DFT could be formulated through an analytic treatment
of the poles of the dynamic polarizability to look like the RPA equation 
[Eq.~(\ref{eq:wft.10})] thus allowing a rapid implementation of TD-DFT in {\it ab initio} quantum chemistry wave function 
codes.  
Thus this is the method now found in nearly every internationally important quantum chemistry and quantum physics code.
The derivation of Ref.~\cite{C95} did not make the adiabatic approximation and allowed for (fixed) fractional occupation numbers.
The matrices ${\bf A}$ and ${\bf B}$ are $\omega$-dependent in the general theory and the number of solutions
exceeds the dimensionality of the matrix problem.  However, for brevity, we will only write the matrix elements 
of the ${\bf A}$ and ${\bf B}$ matrices here in the adiabatic approximation,
\begin{eqnarray}
  A_{ia\sigma,jb\tau} & = & \delta_{\sigma,\tau} \delta_{i,j} \delta_{a,b} (\epsilon_{a\sigma} - \epsilon_{i\sigma})
                   + (ia \vert f_H + f_{xc}^{\sigma,\tau} \vert bj) \nonumber \\
  B_{ia\sigma,bj\tau} & = & (ia \vert f_H + f_{xc}^{\sigma,\tau} \vert jb) \, .
  \label{eq:tddft.13}
\end{eqnarray}

Once LR-TD-DFT equations are formulated for TD-DFT, it is natural to use the TDA and this approximation was 
introduced into the 
TD-DFT literature by Hirata and Head-Gordon \cite{HH99}.  Just as the variational nature of CIS is helpful in circumventing,
or at least, reducing problems associated with triplet and near singlet instabilities in the HF RPA, so too the TDA 
is highly useful when calculating TD-DFT excited-state potential energy surfaces \cite{CGG+00,CJI+07,TTR+08}.
However one should be aware that the cost of the TDA is a loss of oscillator strength sum rules \cite{JCS96} with 
possible implications for the accuracy of corresponding spectra \cite{GMG09}.

We would now like to say a few words about why TD-DFT works from a physical point of view by looking at the
TOTEM.  At least on the lower rungs of Jacob's ladder (LDA and GGAs) the Kohn-Sham orbitals are preprepared 
to describe electronic excitations since the occupied and unoccupied orbitals ``see'' the same xc-potential 
and hence the same number of electrons.  This means that, except for charge transfer excitations which have
large amounts of density relaxation, The TOTEM often applies fairly well in TD-DFT. The corresponding TDA TD-DFT 
singlet and triplet excitation energies are given in {\bf Table~\ref{tab:TOTEM}}.
An examination of the size of the various integrals then shows that, 
\begin{equation}
    \omega_T < \epsilon_a - \epsilon_i < \omega_S \, ,
    \label{eq:tddft.14a}
\end{equation}
with all three values becoming identical for Rydberg states where the differential overlap 
$\psi_i({\bf r})\psi_a({\bf r})$ becomes negligible.
At lower energies we should expect to recover the STEX formulae. As shown in {\bf Table~\ref{tab:TOTEM}}, this is 
possible, at least roughly, within a sufficiently accurate model such as the optimized effective potential (OEP) 
model \cite{C08}.  Direct calculation shows that the Koopmans' theorem interpretation of the occupied orbital
energies as the negatives of ionization potentials holds {\it better} than it does for HF. Thus $I_i=\epsilon_i$.
However the energies of the unoccupied orbitals do not represent the negatives of electron affinities because 
the occupied orbitals ``see'' $N-1$ rather than $N$ electrons.  The quantity $A_a(i^{-1})$ is minus the electron
affinity of the orbital $\psi_a$ for the cation with an electron removed from orbital $\psi_i$.  The electron
thus ``sees'' $N-1$  electrons and so $A_a(i^{-1})$ is to a first approximation equal to $\epsilon_a$ of the 
$N$-electron neutral.  However one Hartree and one exchange interaction with the electron in orbital $\psi_i$
have been removed, so we now have to update the approximation to be 
$A_a(i^{-1}) =  \epsilon_a + (ai \vert f_H + f_{xc}^{\alpha,\alpha} \vert ia)$.  
(Note that the Hartree interaction is estimated in TD-DFT by the integral $(ai \vert f_{xc}^{\alpha,\alpha} \vert ia)$.)

\subsubsection{\color{red} Photochemistry}

The popularity of TD-DFT for modeling absorption spectra is a strong incitation to extend TD-DFT to related 
photophenomena shown in {\bf Fig.~\ref{fig:PES}}.  We are immediately confronted with the fact that TD-DFT
provides only excitation energies, $\omega_I = E_I + E_0$, so that excited-state PESs must be calculated
by adding the excitation energy obtained from TD-DFT to the ground state energy obtained by static
ground-state DFT: $E_I^{TD-DFT} \equiv E_0^{DFT} + \omega_I^{TD-DFT}$.  Exploring excited-state PESs
is facilitated by the implementation of analytic gradients for TD-DFT excited states, 
${\bf \nabla} E_I^{TD-DFT} = {\bf \nabla} E_0^{DFT} + {\bf \nabla} \omega_I^{TD-DFT}$, 
in many electronic structure codes. 

The first and easiest of these phenomena that can be treated is fluorescence.  The TD-DFT fluorescence
energy is a TD-DFT singlet excitation energy calculated at the optimized geometry of the singlet 
excited state.  The corresponding Stokes shift is the difference between the absorption and fluorescence
energies, $\omega_{STOKES} = \omega_{ABS} - \omega_{FLUO}$.  

More challenging is to use TD-DFT to investigate possible photochemical reaction mechanisms.
While TD-DFT is now well-established as a part of the photochemical modelers' toolbox, applications to
studying the entire course of a photochemical reaction are relatively new.
The reader is referred to a recent review on the subject by Casida, Natarajan, and Deutsch \cite{CND11}.
Suffice it to say here that it is now increasingly possible to go beyond Ehrenfest dynamics or the simple pathway
method and carry out mixed TD-DFT/classical trajectory surface hopping calculations to get an idea
of possible photochemical reaction mechanisms and reaction branching ratios 
\cite{TTR07,TTR+08,BPP+10,SF10}.

\subsubsection{\color{red} The Challenge of Excitonics}

To judge by the numerous articles which have appeared in the {\it Annual Review of Physical Chemistry},
excitonics, or the study of excitons, is of great and recurring interest to our community.
It would most likely be described differently by a solid-state physicist
and by a physical chemist or biophysicist.  A solid-state physicist would probably begin with the image of
delocalized states with the expected broad absorption spectra.  Sharp spectral features are then often
associated with localized excitations which could propagate throughout a material and eventually
the electrons and holes could separate to do something interesting such as generate electricity or
drive a chemical reaction.  A physical chemist or biophysicist would probably begin by dissecting
an extended system, such as a molecular solid or a biomolecule, into components which are likely to
have localized excitations and then use an excitonic model to try to understand how excitations interact
to lead to spectral shifts or energy transfer.  In the end the concepts are not very different except
that the solid-state physicist is localizing a delocalized phenomenon while the chemist is delocalizing
localized phenomena.  

These two points of view are now frequently coming together in the area of nanoscience
where solids and molecules frequently meet at the nanointerface.  
This is particularly apparent when modeling dye-sensitized solar cells \cite{DP07}.  There is
a level matching problem in these systems where the excited-states of dye molecules must be aligned with
the band structure of transparent semiconductor particles so, for example, that electrons can pass into the 
conduction band when neutral excitations break up at the molecule-crystal interface.  Calculations
such as that of Ref.~\cite{RGD+09} give an idea of what is currently possible when modeling dye-sensitized 
solar cells with TD-DFT.

Excitons are also highly relevant in photobiology where exitations in nearby proteins are coupled, thus allowing
energy to be captured by antenna systems and transported to active sites where electronic energy is used to
do biologically-important chemistry.  Both for practical and conceptual reasons, it makes sense to use
subsystem density-functional theory.  Johannes Neugebauer has been particularly successful in this regard
in implementing and improving the subsystem version of TD-DFT 
\cite{N07,N08}
initially proposed by Casida and Weso{\l}owski
\cite{CW04}.  A recent application of this theory is to the light harvesting complex II of green
plants \cite{KN11}. 

Given the system sizes involved and the importance of also simulating exciton dynamics, it is probable that 
the TD-DFT will have to be combined with semi-empirical models, such as time-dependent DFTB \cite{N09}, at some
point in order to describe the full complexity of excitonic phenomena.

\subsection{\color{red} Problems, Problem Detection, and Solutions}
\label{sec:tddft_problems}

By discussing the applications of TD-DFT before discussing the problems of TD-DFT, we may have given the
false impression that classical TD-DFT is a blackbox method.  It is far from that!  Just as pure Kohn-Sham
DFT looks like the HF (or rather the Hartree) approximation but behaves very differently, TD-DFT even if put 
in the form of the classic RPA equation of quantum chemistry does not behave like TD-HF.
Conventional LR-TD-DFT in its basic AA using pure DFAs often, 
but not always works better than LR-TD-HF.  It has been found to work best 
for (i) low-energy (ii)  one-electron excitations involving (iii) little or no charge transfer and (iv) which 
are not too delocalized.  This section examines the reasons for these limitations 
and gives a few suggestions as to what can be done when a 
better solution is needed.  Interestingly many of the problems are not due to the adiabatic approximation so 
that our analysis may be carried out in terms of Jacob's jungle gym (Fig.~\ref{fig:junglegym}).
Discussion of problems associated with the adiabatic approximation and possible solutions will be delayed 
till sections~\ref{sec:beyond} and \ref{sec:around}.

\subsubsection{\color{red} Problems associated with $E_{xc}$}

There is an intimate relationship between the ground- and excited-electronic states in conventional TD-DFT.
Not only are the excited-state PESs obtained by adding TD-DFT excitation energies to the DFT ground-state energy,
but the same ${\bf A}$ and ${\bf B}$ matrices that occur in LR-TD-DFT also occur in a stability analysis of 
the DFT ground state \cite{BA96a,CGG+00,C02}.  In fact the theory closely parallels what happens in HF stabililty 
analysis (Sec.~\ref{subsec:stability}).

In order to understand the similarities and differences between what happens in DFT and HF, let us first look
at the classic example of H$_2$.  In Hartree-Fock theory, stretching the bond beyond the Coulson-Fischer point
leads to symmetry breaking in the sense that a DODS solution
becomes lower in energy than a SODS.  The reason is well understood in 
wave-function theory and has to do with presence of ionic contributions in the DODS HF wave function.  
The situation is different in DFT.  Assuming NVR, we must have a SODS solution.  There is thus a unique 
nodeless orbital, $\psi({\bf r})$, which is found by taking the square root of the density.  The noninteracting 
potential is then calculated as, $v_s({\bf r}) = \left( \nabla^2 \psi({\bf r}) / 2 \psi({\bf r}) \right) - \epsilon$, 
with the orbital energy, $\epsilon$, adjusted so that $v_s$ goes to zero at infinity.  The NVR assumption may then 
be explicitly verified by solving the Kohn-Sham equation to make sure that $\epsilon$ does indeed 
correspond to the lowest orbital energy of the noninteracting system.  In this way, it is seen that symmetry breaking 
does not occur in the exact theory.  

But it does occur for DFAs even though it is not as severe as in HF.  In particular 
the Coulson-Fischer point for pure density functionals occurs at larger H$_2$ bond distances than in HF, but it does occur
and is recognizable because the square of the lowest triplet excitation energy, $\omega_T^2$, goes to zero
exactly at the Coulson-Fischer point before becoming negative (an imaginary triplet excitation energy) at
larger distances \cite{CGG+00}.  Incidently it is perfectly possible for an SCF calculation beginning with a
SODS guess to converge to a SODS solution even when there is a lower-energy DODS solution \cite{CIC06}.
In this sense, a TD-DFT calculation may serve as an important check on the correctness of the ground-state
calculation.  However when the goal is to calculate excited-state PESs, the presence of triplet and
near singlet instabilities is highly inconvenient.  As in the HF case, a partial solution is then provided
by the TDA whose excitation energies ``inherit'' some of the variational nature of CIS and so do not
collapse when the ground-state solution becomes unstable \cite{CGG+00,CJI+07,TTR+08}.

A second phenomenon which can occur when breaking a bond is the occurance of a cusp in the ground-state
PES when orbital fillings suddenly change \cite{CJI+07,HNI+10}.  The wave-function solution to this problem
is to allow configurations to mix and so have the wave function be a linear combination of determinants
corresponding to the two different orbital fillings.  In contrast this usually corresponds in DFT calculations
to the situation of effective failure of NVR when the LUMO falls lower in energy than the HOMO.  Since
most DFT programs enforce {\it Aufbau} filling of the orbitals, the LUMO is immediately filled so that it becomes
the next HOMO and the old HOMO the new LUMO.  However the LUMO remains lower than the HOMO and the calculation
simply fails to converge \cite{CJI+07,TTR+08}.  As already discussed in Sec.~\ref{sec:dft_formalism}, this
``strange'' orbital filling means that we are, in fact, in the presence of a spin-wave instability.  The 
conventional DFT solution is to go to ensemble theory and allow the orbitals to have fractional occupation 
number. 
Unfortunately it is really far from clear how well conventional DFAs apply to the ensemble situation.

\marginpar{\hrule $\,$ \\ {\color{blue} RDMFT}:  Reduced density-matrix functional theory. \\ \hrule}
This situation is probably one of the contributing factors for the development of reduced density-matrix functional theory
(RDMFT).  In principle, the density is replaced with the 1-RDM.  However in practice it is the natural orbitals
and their corresponding occupation numbers which become the fundamental variables.  As such it is very much like
ensemble density functional theory but also has the possibility of handling nonlocal potentials (i.e., one-electron
operators which are not simply multiplicative functions). 
Exact equations may also be derived for two-electron systems since L\"owdin and Shull showed that the 2-electron wave function 
may be expanded exactly \cite{LS56}.  The spatial part is given by,
\begin{equation}
  \Psi({\bf r}_1,{\bf r}_2) = \sum_i \sqrt{\frac{n_i}{2}} e^{i2\alpha_i} \phi_i({\bf r}_1) \phi_i({\bf r}_2) \, ,
  \label{eq:tddft.15}
\end{equation}
where the natural orbitals, $\phi_i({\bf r})$, are assumed to be real, their complex phase, $\alpha_i$, has been
exhibited explicitly, and the associated natural orbital occupation numbers are denoted $n_i$.  Defining the
phase is the major problem in the two-electron theory, but typically the minority natural orbitals are found 
empirically to have a negative phase with respect to the dominant phase. 
Very good exact functionals may
then be developed for the two-electron case which, for example, give the correct dissociation of H$_2$. 
Notable early work attempting to develop 1-RDM functionals for systems with more than two electrons include that of 
M\"uller \cite{M84} and Goedecker and Umrigar \cite{GU98}.  For more recent work see, for example, Ref.~\cite{LSD+09} and references therein.

Finally there is a problem with conical intersections which arises precisely because excited-state PESs are calculated 
in TD-DFT as, $E_I=E_0+\omega_I$.  For a molecule with $N_f$ internal degrees of freedom, ${\bf Q}$, the PES is an $N_f$-dimensional
object in an $(N_f+1)$-dimensional space (because we have added the energy axis).  When two PESs cross, we have the
constraints (i) $E_I({\bf Q}) = E_J({\bf Q})$ and (ii) $H_{I,J}({\bf Q})=0$.  Constraint (i) is always present in TD-DFT
and reduces the dimensionality of the crossing space to a hyperline always.  Constraint (ii) should further reduce the
dimensionality of the intersection space to a hyperpoint --- i.e., a conical intersection where there are two independent
directions going away from the intersection space in hyperspace where the surfaces separate (unless of course we have a diatomic
so that $N_f=1$ in which case there is an avoided crossing instead.)  However Levine, Ko, Queenville, and Martinez
have pointed out \cite{LKQM06} that constraint (ii) is absent for interactions between the excited states and the ground
state because of the use of the formula $E_I=E_0+\omega_I$.  We are thus left with the condition that conical intersections
cannot exist in TD-DFT.  While this is certainly correct, in practice they may be approached closely enough to make
mixed TD-DFT/classical surface hopping photodynamics calculations at least qualitatively useful for investigating 
photochemical reaction mechanisms \cite{TTR+08}.

\subsubsection{\color{red} Problems associated with $v_{xc}$}

The problems arising from $E_{xc}$ typically arise at molecular geometries corresponding to bond making or breaking in the
ground state.  Problems associated with $v_{xc}$ occur even in the Franck-Condon region near the ground-state equilibrium
geometry.  In Sec.~\ref{sec:tddft_classic}, we argued that the Kohn-Sham orbital energy difference should lie between the 
corresponding singlet and triplet excitation energies [Eq.~(\ref{eq:tddft.14a})] and that all 
three quantities will be well approximated by the Kohn-Sham orbital energy difference for Rydberg-type excitations.  Since
the limit of a given Rydberg series is just the ionization potential of an orbital $\psi_i$, we see immediately that the
TD-DFT ionization threshold is at minus the HOMO energy.  However practical DFAs have
minus HOMO energies which are typically several electron volts too low compared to the experimental ionization potential
This means that finite basis TD-DFT calculations lead to a collapse of the higher excited states in the region between
minus the HOMO energy and the true ionization potential \cite{CJCS98}.  The reader needs to be keenly aware of this point
because the collapse may not be obvious when only medium-sized basis sets are used, though it will occur once large-enough
basis sets are employed.

One approach that could be taken is simply to ignore the problem.  After all, one might think that that the oscillator strength
distribution should be approximately correct even above the TDLDA ionization threshold at $-\epsilon_{HOMO}$ \cite{WMB03}.
Of course, transitions to bound states will no longer be sharp, making assignments more difficult,
but the spectral features will often still be present.

Probably a better approach is to correct the asymptotic behavior of $v_{xc}$ by a shift-and-splice approach 
(see Ref.~\cite{HZA+03} and references therein.)
This is sufficient, for example, to obtain qualitatively reasonable Rydberg
PESs and hence to see valence-Rydberg avoided crossings in TD-DFT calculations for formaldehyde (H$_2$C=O) \cite{CCS98}.
An even more sophisticated approach is the previously-mentioned OEP model.

A subtler issue which is only occasionally discussed in the DFT literature is that the relative orbital energies of occupied
orbital energies obtained with DFAs may not be correct compared with the orbital energies which should come out of exact
Kohn-Sham DFT calculations.  In Refs.~\cite{C02,HCS02} OEP calculations were used to show that errors in 
ethylene (H$_2$C=CH$_2$) asymptotically-corrected LDA TD-DFT excitation energies were due to small
relative errors in the occupied $\sigma$ and $\pi$ MO energies.

\marginpar{\hrule $\,$ \\ {\color{blue} RSH}: Range-separated hybrid. \\ \hrule}
A final issue which is far from minor is the problem of charge transfer excitations \cite{DWH03}.  Up to this point we have argued that
DFT orbitals are preprepared to describe excitations, but a better statement is that they are preprepared to describe {\it 
neutral} excitations.  Indeed, as discussed just after Eq.~(\ref{eq:wft.24}) HF theory seems to be very well designed for
describing long-range charge transfer.  An inspection of the TDA-TD-TDDFT formulae in {\bf Table~\ref{tab:TOTEM}} 
shows that the long-range charge transfer
energy in TD-DFT is expected to be just the orbital energy difference, $\epsilon_a-\epsilon_i$, which is badly underestimated
because of problems with the asymptotic behavior of $v_{xc}$ related to the PNDD.   This also suggests that the overlap between
the initial and final orbitals in an excitation could be used as a diagnostic tool to anticipate when TD-DFT will break-down due
to charge transfer excitations \cite{PBHT08}.  As the linearized $\Delta$SCF DFT formulae in {\bf Table~\ref{tab:TOTEM}} show, this problem
is not as severe in the $\Delta$SCF approach which takes partial account of orbital relaxation energies and this has been the
basis of one proposal for a charge transfer correction \cite{CGG+00}.  That paper also makes the point that not every charge
transfer excitation is problematic in TD-DFT --- only the ones involving density relaxation.  In fact, Mulliken's prototypical
$^1\Sigma_u$ charge transfer excitation in H$_2$ is an excellent example where density relaxtion is small and TD-DFT has little 
problem.  However a better answer can be found in the range-separated
hybrid (RSH) approach (see e.g., Ref.~\cite{SKB09} and references therein.)
The basic idea here is to separate the electron repulsion, $1/r_{12}$, into a short-ranged part,
$\left( 1/r_{12} \right)_{SR} = \mbox{erfc}(\gamma r_{12})/r_{12}$, which is described by a pure DFA and a long-range part,
$\left( 1/r_{12} \right)_{LR} = \mbox{erf}(\gamma r_{12})/r_{12}$, which is described by a suitable wave-function theory such as
Hartree-Fock.  There are several variations on the name and precise functional forms used in this theory, but they do yield
a dramatic reduction in errors due to the underestimation of charge transfer excitations.  Drawbacks are the presence 
of a possibly system-dependent parameter ($\gamma$ in the above example)

Interestingly a recent OEP study shows that errors in the charge transfer excitation of HeH$^+$ are not due to errors in the
xc-kernel, but rather to errors in Kohn-Sham orbital energies arising from DFAs \cite{GIHG09}.  A second OEP study confirms that a
significant part of the errors in the adiabatic TD-DFT is due to the use of DFAs, but goes on to show that an important residual
error is due to neglet of the frequency dependence of the xc-kernel \cite{HIG09}.

\subsubsection{\color{red} Problems associated with $f_{xc}$}

There are two main problems associated with $f_{xc}$.  The first has to do with the need to include not just 1p1h
excitations but also higher excitations in order to treat some problems.  For example explicit 2p2h excitations are
needed to describe the first singlet excited state of butadiene, CH$_2$=CHCH=CH$_2$, has significant double excitation character~\cite{MTM08}.
Also explicit higher-electron excitations are needed for a proper description of excitations of molecules with open-shell ground states~\cite{C05}.
Although exact TD-DFT should describe this correctly, making the adiabatic approximation restricts the number of solutions of the
LR-TD-DFT equation to the dimensionality of the matrix, namely only the number of 1p1h excitations and de-excitations.  Including some
frequency-dependence in $f_{xc}$ could allow a nonlinear feedback mechanism allowing the matrix equation to have additional solutions.
At one point, it was hoped that going beyond LR to look at, say,
{\it simultaneous} two-photon absorption, would allow two-electron excitations to be treated within the adiabatic
approximation~\cite{GDP96}, however it is now clear that this is not the case.  In particular, the poles of the
dynamic second hyperpolarizability are identical to the poles of the dynamic polarizability~\cite{CT03}, which is to say
the one-electron excitations of adiabatic LR-TD-DFT.  Interestingly two-electron excitations are accessible via {\it sequential} 
absorption in real-time TDHF and TD-DFT~\cite{IL07}.

There is a second problem associated with $f_{xc}$, but also with $v_{xc}$ and even $E_{xc}$.  This problem goes by different names in the literature.
We will refer to it as the ``scale-up catastrophe,'' because it arises when a system is made larger and larger.  Now, an important goal of 
DFT is to extrapolate {\it ab initio} accuracy to molecules too large to otherwise
treat by conventional (i.e., HF-based) {\it ab initio} wave-function calculations.  The usual recommendation is to
test DFT calculations on small molecules where comparisons can be made with accurate {\it ab initio} calculations
and then to {\it assume} that DFT calculations which are reliable for a given class of molecule and molecular
properties will remain accurate as the size of the molecules treated is increased (i.e., scaled-up).
When instead inaccuracies {\it increase} as the size of the system is increased, then we have a scale-up catastrophe.
A perhaps little-known example of a scale-up catastrophe in LR-TD-DFT occurs in periodic systems where the xc-kernel vanishes on the lower
rungs of Jacob's ladder due to improper scaling with respect to the number of $K$ points.~\cite{HHB99}
The result is that the first vertical excitation energy reduces to just the difference between the HOMO and LUMO orbital energies, which is too low.
Nevertheless the spectrum may remain essentially correct because the oscillator strength of this transition may be negligeable.
The better known and very disturbing example of a scale-up catastrophe is 
that studies on oligomers of conducting polymers showed that dynamic polarizabilities could be overestimated by
arbitrarily large amounts by simply going to a large enough oligomer~\cite{CPG+98,CPJ+00}. \cite{CPG+98,CPG+99,CPJ+00}
There are different ways to try to understand this error.  For example, most DFAs are based upon the homogeneous
electron gas which is a model for a perfect conductor.  Applying an electric field to a conductor leads to
the build up of surface charges which cancel the applied fields in the interior of the conductor (Faraday effect).
If the system is behaving too much like a conductor, then the HOMO-LUMO gap is too small due, for example, to an
improper description of the PNDD.  That is an explanation mainly in terms of orbital energies and hence $v_{xc}$.
However the reaction field which cancels the applied field in the interior of the conductor is given by the response of the self-consistent field which
depends on the description of the xc-kernel.  This xc-kernel must be ``ultra-nonlocal'' in the sense that
it must react to charges built up on the surface of the conductor.  On the lower rungs of Jacob's ladder, $f_{xc}^{\sigma_1,\sigma_2}({\bf r}_1,{\bf r}_2)$,
is roughly diagonal in $({\bf r}_1,{\bf r}_2)$.  This is in contradiction with a rough estimate, exact in the case of
two-electron systems, of the exchange-only part of the xc-kernel which suggests that it be more like $-\vert \gamma({\bf r}_1,{\bf r}_2) \vert^2/
\left( \rho({\bf r}_1) r_{12} \rho({\bf r}_2) \right)$ \cite{PGG96}.  This means that a diagonal $f_{xc}$ is certainly incorrect.
There are few choices for dealing with this problem---either one
restricts oneself to medium-sized systems before the onset of the scale-up catastrophe or one climbs Jacob's
ladder to use more sophisticated functionals.  One way to restrict oneself to medium-sized
systems is to use a subsystem theory such as the one used to treat excitons in light harvesting complex II, so that TD-DFT is only
restricted to medium-sized molecules at any given moment.
Climbing Jacob's ladder to functionals with contributions from Hartree-Fock exchange is also helpful as is the use 
of RSHs~\cite{TTY+04,TTH06,VS06,PTS+07}. Another approach which is helpful in some cases is time-dependent
current-density functional with the Vignale-Kohn functional~\cite{QCV03,F06}.  This is because the current
can carry information about perturbations at a distance.

\section{{\color{red} GOING BEYOND THE ADIABATIC APPROXIMATION}}
\label{sec:beyond}

The exact TD-DFT potential [Eq.~(\ref{eq:tddft.9})] is a functional of
the whole history of past densities and the initial conditions.  These
memory effects, are completely neglected in common TD-DFT applications 
using the AA. In fact, the AA assumes that the xc-potential reacts instantaneously
and without memory and so may be expressed in terms of the ordinary DFT
xc-energy evaluated for the density at that time,
\begin{equation}
  v_{xc}^\sigma({\bf r},t) = \frac{\delta E_{xc}[\rho_\alpha^t,\rho_\beta^t]}
       {\delta \rho_\sigma^t({\bf r})} \, .
  \label{eq:beyond.AA}
\end{equation}
Thus the AA xc-potential is only a functional of the function $\rho_\sigma^t({\bf r})$
of the three independent variables ${\bf r} = (x,y,z)$ at fixed time, and not a functional of 
the full function $\rho_\sigma({\bf r},t)$ of four independent variables, a fact that
we have tried to emphasize by our choice of notation.
In this sense, TD-DFAs using the AA work
fairly well when the non-interacting system is a reasonable physical
first approximation to the interacting system, but fail
dramatically in cases like 2p2h excited states \cite{CZMB04}, conical
intersections \cite{LKQM06,CJI+07}, charge transfers and Rydberg
excitations \cite{DWH03,HIG09}, excited states along the bond
dissociation coordinate \cite{GB08}, one-dimensional extended systems
\cite{FBL+02,VMR08}, band-gaps in solids \cite{ORR02} and spectra of
semiconductors \cite{ORR02,RORO02,SAOR03}. In such cases, the
contribution of past densities to the xc energy becomes essential,
requiring memory functionals specifically designed for TD-DFT. In this
section, we will review the most successful memory functionals
available at the present time. 
Few attempts have tried to approximate directly the
xc-action \cite{KB04}.  Instead, most memory functionals approximate
the derivatives.  Therefore, our discussion will focus on the
construction of memory xc-potentials and specially memory xc-kernels.


Memory introduces strong requirements for TD-DFT functionals
\cite{WB11}, which forbids the intuitively simple functional
approximations of ground-state DFT. For example, a simple
parameterization of a spatially-local memory LDA xc-kernel for the
homogeneous electron gas  \cite{GK85} violates, by construction,
the zero-force theorem \cite{D94,V06}, introducing unphysical damping
effects during the time propagation of the density and a corresponding
lose of total energy. Several schemes have appeared that enforce the
sum rules \textit{a posteriori} \cite{DBG97,KB08}. Most importantly,
these studies showed that memory functionals must be non-local in
space. This property forbids the application of the popular gradient
expansion to memory functionals \cite{VK96}, ultimately a consequence
of the ultra-nonlocality property of the exact xc-functional.

\subsection{\color{red} Current and Lagrangian Density-Functional Theory}
\label{sec:beyond_current}

The ultra-nonlocality property of the exact functional indicates that
the density is perhaps not the best reduced variable to express
time-dependent xc-effects. This redirected the attention to other
time-dependent theories where memory effects are described by local
functionals, such as in TD-CDFT \cite{VK96,VUC97,UV02,TP03,KB04}, in
Lagrangian TD-DFT  \cite{T05a,T05b,UT06}.

TD-CDFT is very much related to TD-DFT, though its range of applications
is broader. In fact, the one-to-one correspondence between currents
and external potentials is the first step in the Runge-Gross existence
proof for TD-DFT \cite{RG84}. However, the TD-CDFT proof does not resort
to any surface integral, which makes its application to solids more
straightforward than TD-DFT. The first memory functional for TD-CDFT is
due to Vignale and Kohn, which consists of a parameterization of a
memory LDA xc-kernel tensor for the homogeneous electron gas \cite{VK96}. This kernel is
local in space and, unlike the corresponding in TD-DFT, it satisfies
the zero-force theorem. However, the kernel introduces too strong
damping effects that induce a decoherence mechanism, which allows the
density to decay back to a ground-state with higher entropy, different
from the real ground-state, in which the total energy is conserved
\cite{DV06}.  The Vignale-Kohn kernel has been used successfully to
calculate the static polarizabilities of one-dimensional conjugated
chains, which were largely overestimated by common adiabatic LDA
\cite{FBL+02}. However, it fails for one-dimensional hydrogen chains,
requiring more sophisticated kernels \cite{VMR08}.  The Vignale-Kohn
kernel also shows the characteristic double-peak spectrum of bulk
silicon \cite{BKB+01}, a feature that is not present in the adiabatic
LDA spectrum. Memory kernels of TD-CDFT can then be map into TD-DFT
kernels maintaining part of the features of the Vignale-Kohn kernel
\cite{NVC10}.


Lagrangian TD-DFT has less applications to realistic systems, but it has
provided deeper insight on the role of memory. Lagrangian TD-DFT is a
reformulation of TD-DFT, in which the electronic density is re-expressed
in a reference frame (Lagrangian frame) that moves along with the
fluid \cite{T05a,T05b}. This allows the memory to be expressed locally in
terms of a the position of fluid elements and the deformation
tensor. The deformation tensor accounts for the Coulomb coupling of a
differential volume of fluid at position ${\bf r}$ and time $t$ with
was at a different position ${\bf r}'$ at an earlier time $t'<t$. Both
TD-CDFT and Lagrangian TD-DFT are equivalent in the linear regime \cite{UT06}.

\subsection{\color{red} Optimized Effective Potential Approaches}
\label{sec:beyond_OEP}

All the aforementioned memory functionals lack a systematic route to
improve upon the present approximations. A more systematic way to
construct memory functionals is to use the TD-OEP,
\marginpar{\hrule $\,$ \\ 
{\color{blue} MBPT}: Many-body perturbation theory.\\ 
 \hrule}
in which many-body perturbation theory (MBPT) quantities are mapped into TD-DFT
\cite{UGG95,G98,L96,HIGB02,BDLS05,BSO+05,HC10}. An easy way to
establish this mapping is to make use of the Kohn-Sham assumption,
i.e., requiring that the interacting time-dependent density equals the
non-interacting one \cite{L96}.  This is the basis of the
time-dependent Sham-Schl{\"uter} equation that allows us to relate
the self-energy to the xc-potential and the kernel of the
Bethe-Salpeter equation to the xc-kernel. In this way, consistent
approximations are possible, that guarantee the satisfaction of exact
conditions \cite{BDLS05}. There are many ways to derive
the time-dependent Sham-Schl{\"u}ter equation, but similar functionals
are obtained if the same MBPT approximation is used. For deriving
xc-kernels, a simple way is to require that the diagonal of the MBPT
response functions leads to the same spectrum as the TD-DFT response
function [Eq.~(\ref{eq:dft.1})], which leads to
\begin{eqnarray}
  \label{eq:beyond.1}
  &&f_{xc}({\bf r},{\bf r}';\omega) = \\
  &&\int{d^3r_1\int{d^3r_2\int{d^3r_3\int{d^3r_4{}}}}}
            \Lambda({\bf r};{\bf r}_1,{\bf r}_2;\omega)
            K({\bf r}_1,{\bf r}_2;{\bf r}_3,{\bf r}_4;\omega)
            \Lambda^\dagger_s({\bf r}_3,{\bf r}_4;{\bf r}';\omega) 
            \, ,\nonumber
\end{eqnarray} 
relating the exact TD-DFT xc-kernel to the kernel $K(\omega)=\Pi_s^{-1}(\omega)-\Pi^{-1}(\omega)$ 
defined from the polarization propagator from MBPT \cite{HC10}.
The mapping is possible thanks to the localization function $\Lambda$, which is expressed in terms of the response
functions [Eq.~(\ref{eq:dft.1})],
\begin{equation}
   \label{eq:beyond.2}
   \Lambda({\bf r}_1;{\bf r}_2,{\bf r}_3;\omega) = 
      \chi^{-1}({\bf r}_1,{\bf r}_2;\omega)
      \chi({\bf r}_2,{\bf r}_3;\omega) \, .
\end{equation} 
The localizer $\Lambda_s$, appearing in Eq.~(\ref{eq:beyond.1}), 
is defined as in Eq.~(\ref{eq:beyond.2}) except in terms of the noninteracting response function ($\chi_s(\omega)$).
The localizer introduces an extra frequency-dependence in the kernel, which brings in the MBPT quantity
the ultra-nonlocality property of the exact TD-DFT functional
\cite{GS99,BSO+05,GOR+07,HC10}. To date, little is known about the
exact role of the localization step, though it is known to have
interesting properties. For example, Gonze and Scheffler showed that
the single-pole approximation when one applies the linear
Sham-Schl{\"u}ter equation in Eq.~(\ref{eq:beyond.2}) gives
\cite{GS99}
\begin{equation}
  \omega_{ia} = \Delta \epsilon_{ia} + (ia|f_{xc}(\Delta\epsilon_{ia})|ai)
              = \Delta \epsilon_{ia} + (ii|K(\Delta\epsilon_{ia})|aa) \, ,
  \label{eq:beyond.3}
\end{equation}
where $\Delta\epsilon_{ia} = \epsilon_a - \epsilon_i$. In fact, a
cancellation of the localizer is observed, and the exact xc-kernel is
equal to a different matrix element of the kernel derived from MBPT
quantities.  A general treatment of localization is plagued with
numerical problems, due to the difficulties in inverting the singular
density response functions that are involved \cite{MK87,HB09}. In the
linear-response regime, an interesting work around can be used to
fully account for the localization effects without the need of
explicitly inverting any response function. This is achieved by
re-expressing LR-TD-DFT as the response of the non-interacting potential
\cite{HG11}.

Using a first-order approximation of $K$, one obtains the TD-EXX
kernel \cite{G98,HIGB02,BDLS05,HC10}. This
kernel provides correct description of charge transfer \cite{HIG09}
and correct position of the continuum states \cite{BKK11} due to the
correct asymptotic decay of exchange. Although the exchange kernel in
MBPT is frequency independent, the full TD-EXX is a frequency-dependent
functional \cite{G98,HB09} due to the localizer $\Lambda$. Therefore
some (minor) memory effects are taken into account \cite{WU08}. In
general, the frequency dependence of the localizer in the TD-EXX is not
enough to include 2p2h- and higher-ph excited poles
\cite{HFT+11,EGCM11}, and frequency-dependent MBPT kernels are
required \cite{C05,RSB+09,HC10,SRO+11}.


The lack of 2p2h excitations is an endemic problem of the adiabatic
approximation \cite{BKK11}. These poles should be present, since TD-DFT
is formally exact. However, they are not present when an adiabatic
xc-kernel is used in the LR-TD-DFT equations. Clearly, the number of
solutions of Eq.~(\ref{eq:tddft.13}) is given by the dimension of the
matrices, that is, the number of 1p1h states. Recovering of 2p2h
states require a frequency-dependent xc-kernels \cite{MZCB04,CZMB04},
which make the LR-TD-DFT equations non-linear and extra solutions
appear. Frequency dependence can be included via the solution of a
Scham-Schl{\"u}ter equation.  The MBPT kernel of the Bethe-Salpeter
equation has to be constructed carefully, since otherwise the
resulting xc-kernel might introduce spurious transitions
\cite{RSB+09,SRO+11}. 

The first successful $\omega$-dependent
xc-kernel is the kernel of dressed TD-DFT, first proposed by
Burke, Maitra and coworkers \cite{MZCB04,CZMB04} to include one 2p2h
at a time, and later generalized to any number of 2p2h states by using
full-featured MBPT techniques \cite{C05,HC10}. The kernel of dressed TD-DFT
can be viewed as a hybrid kernel between the adiabatic xc-kernel and
the frequency-dependent part of the MBPT kernel,
\begin{equation}
  f_{xc}({\bf r},{\bf r'};\omega) = f^{AA}_{xc}({\bf r},{\bf r}') 
  + K^\omega({\bf r},{\bf r};{\bf r}',{\bf r}';\omega) \, ,
  \label{eq:beyond.4}
\end{equation}
where $K^\omega({\bf r},{\bf r};{\bf r}',{\bf r}';\omega)$ corresponds
to the frequency-dependent part of the kernel derived from a
second-order polarization propagator approach (for the mathematical
expression see \cite{HC10} and \cite{BKK11}.) 
Several variations on dressed TD-DFT are possible.  However Ref.~\cite{BKK11} explores
quite a few of these and comparison of the performance of dressed TD-DFT against
benchmark \textit{ab initio} results for 28 organic chromophores goes far
towards establishing the best protocol to use.


\subsection{\color{red} Time-Dependent Reduced Density Matrix Functional Theory}
\label{sec:around_RDMFT}

Important recent work has been done by Giesbertz and coworkers extending RDMFT to the time-dependent case
\cite{PGGB07,GB08,GBG08,GPGB09,GGB10a,GGB10b,GB10}.  
Oddly enough most of these attempts fail to obtain the correct $\omega \rightarrow 0$ limit of the adiabatic approximation
for TD-RDMFT even for two-electron systems \cite{GGB10b}.
However including not only the orbitals and occupation numbers in the time-dependent response equations but also their {\it phases}
allows exact solutions to be obtained for two-electron systems \cite{GGB10a}.  
The observation of the importance of phases is reminiscent of our observations regarding spin-wave instabilities and
violation of NVR and might be taken as further proof of the need to include phase information TD-DFT.
Note that this phase information is not included in the 1-RDM, so that a new functional dependence is needed which goes 
beyond the density-matrix alone.  It may perhaps be assimilated with a memory effect.


\section{{\color{red} GOING AROUND THE ADIABATIC APPROXIMATION}}
\label{sec:around}

In the last section we described progress in TD-DFT going beyond the adiabatic approximation
but trying to keep the essential spirit of TD-DFT.  In this section we describe approaches
which try to solve the problems of the TD-DFT adiabatic approximation while still retaining
something resembling frequency-independent kernel.  They may be thought of as ways to go
{\it around} the problem.

\subsection{\color{red} Reconciling TD-DFT and $\Delta$SCF}
\label{sec:around_reconcile}

We have seen that the $\Delta$SCF method and its MSM variant predate TD-DFT in quantum chemistry.  The older
method is particularly well justified for calculating the first ionization potentials and electron affinities,
the lowest triplet excitation, and perhaps also transitions in general to states dominated by a single configuration.
Indeed the numbers produced by the MSM $\Delta$SCF method and TD-DFT are often quite similar, though the formulae
are different.  It has been a puzzle since the beginning of applications of TD-DFT in quantum chemistry \cite{C95}
to try to understand why this is so.  Reconciling the two methods might also allow some of the better aspects of
the MSM $\Delta$SCF method to be used in TD-DFT, namely the ease of handling 2p2h and higher-ph excitations in
the $\Delta$SCF method or apparently better ability to describe density-relaxation during excitation.  A first
step in this direction were presented by one of us in 1999 and may be found in the on-line proceedings \cite{C99b}.
These ideas were later used to develop a charge-transfer correction for TD-DFT \cite{CGG+00}.  
Hu, Sugino, and Miyamoto  have followed up this work to develop their
own approach to correcting problems with underestimated Rydberg and charge transfer
excitations.~\cite{HSM06,HS07}

\marginpar{\hrule $\,$\\
{\color{blue} CV}: constrained variational \\
\hrule}
More recently these ideas have been very nicely followed up in Tom Ziegler's group in Alberta in the form of 
constrained variational DFT (CV-DFT) \cite{ZSKA08,ZSK+09a,ZSK+09,ZM10}.  The energy is to be minimized with respect
to the transformation, $\phi_i^\prime = \phi_i + \sum_a U_{a,i} \phi_a + {\cal O}^{(2)}(U)$,
of the Kohn-Sham orbitals,
subject to the constraint $\vec{U}^\dagger \vec{U}=1$.  The second-order CV(2)-DFT energy expression
is already familiar from stability analysis (Eq.~(\ref{eq:wft.25}) with $\lambda=1$).  Minimizing to this order just gives the TD-DFT TDA.  However fourth order CV(4)-DFT with relaxed orbitals
gives a much improved treatment of charge transfer excitations with out inclusion of any nonadiabatic frequency dependence \cite{ZK10}.

\subsection{\color{red} Spin-Flip TD-DFT}
\label{sec:around_spinflip}

\begin{figure}
  \begin{center}
    \includegraphics[angle=0,width=1.0\textwidth]{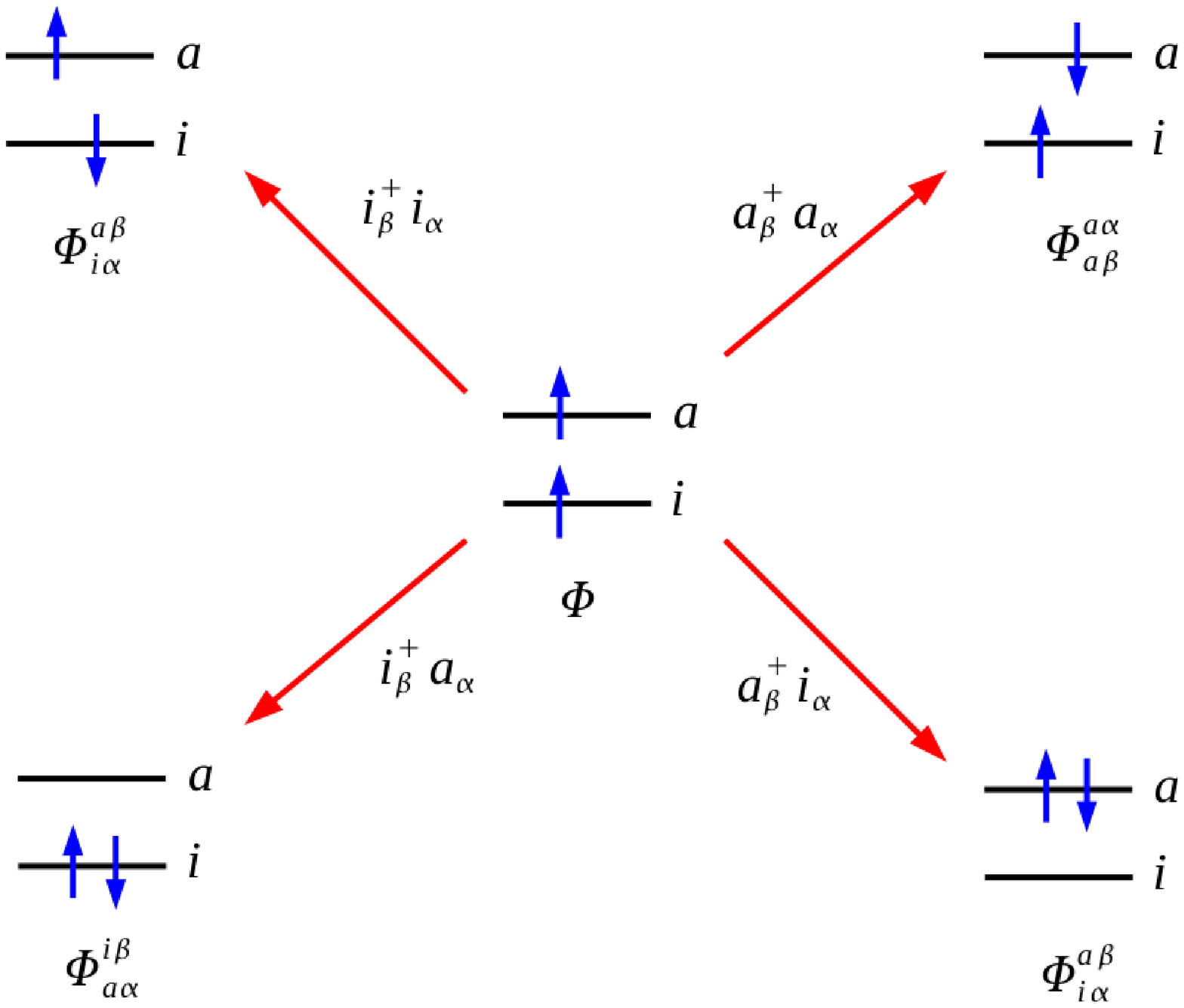}
  \end{center}
  \caption{\label{fig:SFTOTEM}
    Schematic of spin-flip excitations in the two-orbital two-electron model.}
\end{figure}

Another way to go around the adiabatic approximation---this time to access a 2p2h excitation---is to use the spin-flip method.
The basic idea is illustrated in the TOTEM in Fig.~\ref{fig:SFTOTEM}.  
To do this properly, we must consider the Kohn-Sham orbitals as spinors, $\left( \psi_\alpha({\bf r}) , \psi_\beta({\bf r}) \right)$,  
which can rotate their spin-direction at different points in physical space.  The density also acquires spin-labels,
\begin{equation}
  {\bf \rho}({\bf r}) = \left[ \begin{array}{cc} \rho_{\alpha,\alpha}({\bf r}) & \rho_{\alpha,\beta}({\bf r}) \\
                \rho_{\beta,\alpha}({\bf r}) & \rho_{\alpha,\alpha}({\bf r}) \end{array} \right] \, ,
  \label{eq:around.2}
\end{equation}
So the xc-kernel, $f_{xc}^{\sigma_1,\sigma_2;\tau_1,\tau_2}({\bf r},{\bf r}')$, also acquires additional labels.  However this
is just for the purposes of deriving a more general LR-TD-DFT equation.  In the end, we will evaluate the xc-kernel using
the usual collinear Kohn-Sham theory (i.e., each spin orbital is either $\alpha$ or $\beta$ but not some linear
combination of the two).  However it turns out that conventional pure DFAs do not allow spin-flips.

In order to get good agreement with experiment, Ana Krylov and coworkers \cite{SK03,SHG03} found it necessary to use a significantly
higher amount of HF exchange (50\%) than is typically used for ground state properties ($\sim$ 25\%).
Even higher percentages of HF exchange ($> 50$\%) have been reported to be necessary for calculating second
hyperpolarizabilities of diradical systems by this spin-flip method~\cite{KNO+07}.
Although the use of a different functional for ground and excited states is disturbing, the basic idea is
admirable and this method continues to be used~\cite{KNO+07,TSP+08,DGP08}.  In particular, this is the spin-flip TD-DFT
approach mentioned in the introduction in the context of its recent use by Minezawa and Gordon who found
the method to give a relatively good description of conical intersections in ethylene~\cite{MG09}.

The next and most recent major advance in spin-flip TD-DFT came with an article by Wang and Ziegler~\cite{WZ04}.
It is intimately related to work by Wenjian Liu and coworkers on relativistic four-component TD-DFT~\cite{GZL+05}.
Wang and Ziegler proposed that any pure spin-density xc-functional, $E_{xc}[\rho_\alpha,\rho_\beta]$, could be used to make a noncollinear 
xc-functional suitable for spin-flip calculations by making the substitution,
\begin{equation}
  \rho_\alpha  \rightarrow  \rho_+ = \frac{1}{2} \left( \rho + s \right)  \, \, \, , \, \, \,
  \rho_\beta   \rightarrow  \rho_- = \frac{1}{2} \left( \rho - s \right) \, ,
  \label{eq:around.3}
\end{equation}
involving two quantities which are invariant under a unitary transformation of the spin coordinates.  These are the
total charge density, $\rho = \rho_{\alpha,\alpha} + \rho_{\beta,\beta}$,
and the magnetization, $s$, whose square is given by, $s^2 = \left( \rho_{\alpha,\alpha} - \rho_{\beta,\beta} \right)^2 + 2 \left( \rho_{\alpha,\beta}^2
+ \rho_{\beta,\alpha}^2 \right)$.  The collinear limit of $s$ is just the spin-polarization, $ s \rightarrow \rho_\alpha - \rho_\beta$,
after an appropriate choice of phase.  The factor of $1/2$ has been introduced by us \cite{HNI+10} so that,
\begin{equation}
  \rho_+  \rightarrow  \rho_\alpha \, \, \, , \, \, \, \rho_-  \rightarrow  \rho_\beta \, ,
  \label{eq:around.4}
\end{equation}
in the same limit.  After taking derivatives and the noncollinear limit, the xc-kernel becomes,
\begin{equation}
  \left[ \begin{array}{cccc} f^{\alpha,\alpha;\alpha,\alpha}_{xc} & f^{\alpha,\alpha;\beta \beta}_{xc}
                   & f^{\alpha,\alpha;\alpha,\beta}_{xc} & f^{\alpha,\alpha;\beta,\alpha}_{xc} \\
                    f^{\beta,\beta;\alpha,\alpha}_{xc} & f^{\beta,\beta;\beta,\beta}_{xc}
                   & f^{\beta,\beta;\alpha,\beta}_{xc} & f^{\beta,\beta;\beta,\alpha}_{xc} \\
                    f^{\alpha,\beta;\alpha,\alpha}_{xc} & f^{\alpha,\beta;\beta,\beta}_{xc}
                   & f^{\alpha,\beta;\alpha,\beta}_{xc} & f^{\alpha,\beta;\beta,\alpha}_{xc} \\
                    f^{\beta,\alpha;\alpha,\alpha}_{xc} & f^{\beta,\alpha;\beta \beta}_{xc}
                   & f^{\beta,\alpha;\alpha,\beta}_{xc} & f^{\beta,\alpha;\beta,\alpha}_{xc} \end{array} \right]
   =
   \left[ \begin{array}{cccc} f^{\alpha,\alpha}_{xc} & f^{\alpha,\beta}_{xc} & 0 & 0 \\
                               f^{\beta,\alpha}_{xc} & f^{\beta,\beta}_{xc} & 0 & 0 \\
                                0 & 0 & \frac{v^{\alpha}_{xc}-v^{\beta}_{xc}}{\rho_\alpha-\rho_\beta} & 0 \\
                                0 & 0 & 0 & \frac{v^{\alpha}_{xc}-v^{\beta}_{xc}}{\rho_\alpha-\rho_\beta} \end{array} \right]
  \, .
  \label{eq:around.5}
\end{equation}
This approach to spin-flip TD-DFT has been applied to the dissociation of H$_2$~\cite{WZ04} and to calculate the spectra
of open-shell molecules~\cite{FW05,SZ05,GWZC06}.  
We have applied the Wang-Ziegler approach to the photochemical ring opening of oxirane (H$_2$COCH$_2$) \cite{HNI+10}.
While this approach definitely overcame the cusp problem for $C_{2v}$ ring opening and does allow coupling of the ground-
and excited-states so that there is a true conical intersection, the position of the conical intersection was found to
be intermediate between the CIS seam and the complete active space SCF conical intersection \cite{HNI+10}.  Very recently analytical derivatives
have been worked out for the Wang-Ziegler method \cite{SMZ11}.

The Wang-Ziegler approach has also been proposed as the basis of a more general spin-coupled TD-DFT~\cite{VR07} and applied \cite{RA10}.
Very recently it has been used to solve the problem of spin-adaption for LR-TD-DFT in open-shell molecules \cite{LL10,LLZS11}
that had been highlighted by earlier work on spin contamination in TD-DFT \cite{ICJC09}.
Hybrid spin-flip functionals are also beginning to appear \cite{RA10,RVA10}.

\section{{\color{red} PERSPECTIVES}}
\label{sec:conclude}

In coming to the final section of this review, we are deeply conscious of the many aspects of
TD-DFT which we have neglected and the shallowness of our treatment of those aspects of TD-DFT
which we have been able to treat in this small overview of what has become a rather vast field.
Nevertheless we hope that the reader has appreciated the historical interplay between DFT, TD-DFT,
and the development of DFAs.  It is a fact that adiabatic TD-DFT in seeking to calculate new types
of properties, has placed brought into sharper focus known problems of and has placed new demands on DFAs. 
This in turn has pushed further improvement of DFAs for classical DFT and pushed more detailed exploration 
of Jacob's ladder.  

Also significant has been the difficulty of answering the Runge-Gross challenge to go beyond
the AA and include memory in TD-DFT.  Section~\ref{sec:beyond} has discussed
methods for trying to go beyond the AA while still trying to preserve the
true spirit of (TD-)DFT --- that is, to do as much work analytically before asking the computer
to do the work.  In the end, this means taking things from MBPT and putting them into the functionals.
When we do the opposite --- that is adding MBPT on top of the functionals --- then that is
an indication of something that we do not yet understand well enough and need to work
on some more.  

The penultimate section (Sec.~\ref{sec:around}) discussed primarily pragmatic ways to try
to circumvent problems created by the AA without abandoning the AA.  These, plus dressed TD-DFT
(Sec.~\ref{sec:beyond}), are the tools we have now but which may merge in the future with the more 
rigorous approaches of Sec.~\ref{sec:beyond} to give something approaching an answer to the Runge-Gross challenge.

Only time will tell $\ldots$ but then time is such an important and fundamental physical parameter!

\subsection*{{\color{blue} Summary Points}}

{\color{blue}
\begin{enumerate}

\item After almost three decades, the Runge-Gross theorem is largely uncontested as the foundations of a rigorous 
      theory.  Initial problems in the definition of the action have been heavily invetigated, leading to 
      increasingly satisfactory definitions of appropriate xc-actions.  Now most fundamental TD-DFT research is devoted
      to improving TD-DFAs.

\item TD-DFT is making good progress towards a general application to photochemical problems. Charge transfer, 
      Rydberg excited states and 2-particle/2-hole states can now be correctly treated within LR-TD-DFT. The 
      treatment of conical intersections will require overcoming the non-interacting $v$-representability problem 
      in DFT, which requires fractional occupation numbers and very possibly complex orbitals.

\item In general, LR-TD-DFT using the Tamm-Dancoff approximation leads to better excitation energies and it should be 
      preferred when calculating potential energy surfaces. However, the Tamm-Dancoff approximation can lead to inaccurate
      oscillator strength distributions.

\item Most of the problems attributed to LR-TD-DFT can be simply resolved by using decent approximations of the DFT 
      functionals. From practical experience, asymptotically-corrected hybrid functionals with 20-25\% of HF exchange 
      lead to the best results for singly-excited states of finite systems.

\item Memory functionals have extended the range of applicability of TD-DFT, though some problems are yet to be 
      resolved. The longstanding problem of 2-particle/2-hole states can be resolved by adding to the adiabatic 
      approximation a frequency-dependent component derived with the help of MBPT techniques. Though the mixture 
      requires careful application to avoid double-counting of correlation, present applications show encouraging results.

\item Spin-Flip LR-TD-DFT currently offers a practical solution for photochemical problems. It adds some double 
      excitation character and it avoids regions of the potential energy surface with noninteracting  
      $v$-representability problems, offering an easy route to calculate conical intersections. Also, it can 
      successfully treat excited states of some open-shell systems.

\end{enumerate}
}

\subsection*{{\color{magenta} Future Issues}}

{\color{magenta}
\begin{enumerate}

\item Even though some advances have been made, memory functionals are still in their infancy, and better functionals 
      are necessary for extending the range of application of TD-DFT. It seems more appropriate to think in terms of 
      a different Jacob's ladder for memory functionals in TD-DFT with the rungs defined according to the functional 
      dependence as: TD-DFT (density), TD-CDFT (density and current) and L-TDDFT (fluid position and deformation tensor), 
      TD-OEP (orbitals), TD-RDMFT (1-electron density matrix, orbital phase).

\item A general application of TD-DFT to photochemistry will require more sophisticated functionals, depending probably 
      on additional parameters other than just the density.  Studies of TD-RDMFT indicate that TD-DFT functionals may
      also need as variables the occupation number and the time-dependent phase of the orbitals.

\end{enumerate}
}

\section*{{\color{red} ACKNOWLEDGEMENTS}}

M.\ H.-R.\ would like to acknowledge a scholarship frm the French Ministry of Education.
This work has been carried out in the context of the French Rh\^one-Alpes
{\it R\'eseau th\'ematique de recherche avanc\'ee (RTRA): Nanosciences aux limites de la nano\'electronique}
and the Rh\^one-Alpes Associated Node of the European Theoretical Spectroscopy Facility (ETSF).
Klaas Giesbertz is thanked for sending us a copy of his doctoral thesis.  
Evert Jan Baerends, Christoph Jacob, Melvin Levy, Johannes Neugebauer, 
Lucia Reining, Gustavo Scuseria, and Tomasz Weso{\l}owski are thanked for helpful conversations.


\bibliographystyle{arnuke_revised}
\bibliography{refs}
\end{document}